%% file: marked.tex
\documentclass[11pt]{article}

\usepackage[utf8]{inputenc}
\usepackage[OT1]{fontenc}

\usepackage{graphicx}

\usepackage{amsmath, amsthm, amssymb}

\theoremstyle{plain}

\newtheorem{proposition}{Proposition}[section]
\newtheorem{theorem}{Theorem}[section]
\theoremstyle{definition}
\newtheorem{example}{Example}
\newtheorem{remark}{Remark}
\numberwithin{equation}{section}
\allowdisplaybreaks
\usepackage{bm}
\usepackage{mathdots}

\usepackage{multirow}
\usepackage{booktabs}
\usepackage{rotating}

\usepackage{color}

\usepackage{fullpage}
\usepackage{setspace}
\usepackage{csquotes}
\MakeOuterQuote{"}

\usepackage{natbib}

\usepackage{authblk}

\usepackage{comment}
\excludecomment{en-text}
\includecomment{jp-text}
\includecomment{comment}

\input nakamacro-markedVersion

\title{Marked point processes and intensity ratios \\ for limit order book modeling}

\author[1]{Ioane Muni Toke\thanks{Bâtiment Bouygues, 3 rue Joliot Curie, 91190 Gif-sur-Yvette, France. ioane.muni-toke@centralesupelec.fr}}
\author[2]{Nakahiro Yoshida\thanks{3-8-1 Komaba, Meguro-ku, Tokyo 153- 8914, Japan. nakahiro@ms.u-tokyo.ac.jp}}
\affil[1]{Université Paris-Saclay, CentraleSupélec, Mathématiques et Informatique pour la Complexité et les Systèmes, France.}
\affil[2]{Graduate School of Mathematical Sciences, University of Tokyo, Japan.}
\affil[1,2]{CREST, Japan Science and Technology Agency, Japan.}

\begin{document}

\maketitle

\begin{abstract}
This paper extends the analysis of \cite{MuniTokeYoshida2019} to the case of marked point processes. We consider multiple marked point processes with intensities defined by three multiplicative components, namely a common baseline intensity, a state-dependent component specific to each process, and a state-dependent component specific to each mark within each process. We show that for specific mark distributions, this model is a combination of the ratio models defined in \cite{MuniTokeYoshida2019}. We prove convergence results for the quasi-maximum and quasi-Bayesian likelihood estimators of this model and provide numerical illustrations of the asymptotic variances. We use these ratio processes in order to model transactions occuring in a limit order book. Model flexibility allows us to investigate both state-dependency (emphasizing the role of imbalance and spread as significant signals) and clustering. Calibration, model selection and prediction results are reported for high-frequency trading data on multiple stocks traded on Euronext Paris. We show that the marked ratio model outperforms other intensity-based methods (such as "pure" Hawkes-based methods) in predicting the sign and aggressiveness of market orders on financial markets.

\halflineskip
\end{abstract}

\vfill
\pagebreak

\section{Introduction}

The limit order book is the central structure that aggregates buy and sell intentions of all the market participants on a given exchange. This structure typically evolves at a very high-frequency: on the Paris Euronext stock exchange, the limit order book of a common stock is modified several hundreds of thousand times per day. Among these changes, thousands or tens of thousand events account for a transaction between two participants. The rest of the events indicate either the intention to buy/sell at a limit price lower/higher than available, or the cancellation of such intentions \citep{Abergel2016}.

Empirical observation of high-frequency events on a limit order book may reveal irregular interval times (durations), clustering, intraday seasonality, etc. \citep{Chakraborti2011}. Stochastic point processes are thus natural candidates for the modeling of such systems and their time series \citep{Hautsch2011}. In particular, Hawkes processes have been successfully suggested for the modeling of limit order book events \citep{Bowsher2007,Large2007,Bacry2012,Bacry2013,MuniTokePomponio2012,
Lallouache2016,LuAbergel2018}.

One drawback of such models is the difficulty to account for high intraday variability. Another drawback of such models is the lack of state-dependency: the observed state of the limit order book does not influence the dynamics of the events. One may try to include state-dependency by specifying a fully parametric model \citep{MuniTokeYoshida2017}, which is a cumbersome solution. Another solution is to extend the Hawkes framework with marks \citep{Rambaldi2017} or with state-dependent kernels \citep{Morariu2018}. \cite{MuniTokeYoshida2019} has shown that state-dependency can be efficiently tackled by a multiplicative model with two components: a shared baseline intensity and a state-dependent process-specific component. An intensity ratio model can then allow for efficient estimation of state-dependency. Several microstructure examples are worked out, including a ratio model for the prediction of the next trade sign\footnote{When characterizing a market order, we use indistinctly the terms \emph{side} (bid/ask) or \emph{sign} (-1,+1) to indicate if a transaction occurs at the best bid or best ask price of the limit order book.}.

In this work, we extend the framework of \cite{MuniTokeYoshida2019} to some cases of marked point processes, by adding a third term to the multiplicative definition of the intensity, which accounts for some mark distribution. We use this extension to deepen our investigation of limit order book data. In financial microstructure, one of the characteristics of an order sent to a financial exchange is its aggressiveness \citep{Biais1995,Harris1996}. We will say here that an order is aggressive if it moves the price. A ratio model with marks can thus be used to analyse both the side (bid or ask) and aggressiveness of market oders.

The rest of the paper is organized as follows. In Section \ref{sec:MarkedProcesses} we show that some marked models can be viewed as combinations of intensity ratios of non-marked processes. Section \ref{sec:QuasiLikelihoodEstimation} defines the quasi-likelihood maximum and Bayesian estimators and proceeeds to the analysis of the estimation. Theorem \ref{thm:ConvergenceResult} states the convergence result and a numerical illustration follows. We then turn to the main financial application in Section \ref{sec:TradeAggressiveness}, and show how the two-step ratio model can efficiently predict (in a theoretical setting) the sign and aggressiveness of the next trade. Finally, the full proof of Theorem \ref{thm:ConvergenceResult} is given in Appendix \ref{sec:TheoremProof}, and for completeness elements on quasi-likelihood analysis are recalled in Appendix \ref{sec:QLArefresher}.

\section{Marked process models as two-step ratio models}
\label{sec:MarkedProcesses}
Let $\bbI=\{0,1,...,\bar{i}\}$. We consider certain marked point processes 
$N^i=(N^i_t)_{t\in\bbR_+}$, $i\in\bbI$ and $\bbR_+=[0,\infty)$. 
For each $i\in\bbI$, let $\bar{k}_i$ be a positive integer, and 
let $\bbK_i=\{0,1,...,\bar{k}_i\}$ be a space of marks for the process $N^i$. 
We denote by $N^{i,k_i}=(N^{i,k_i}_t)_{t\in\bbR_+}$ the process counting events of type $i$ with mark $k_i\in\bbK_i$. 
We have obviously $N^i=\sum_{k_i\in\bbK_i}N^{i,k_i}$. 
Let $\check{\bbI}=\cup_{i\in\bbI}\big(\{i\}\times\bbK_{i}\big)$. 
We assume that the intensity of the process $N^i$ with mark $k_i$, i.e., 
the intensity of $N^{i,k_i}$, is given by  
\beas 
\lambda^{i,k_i}(t,\vartheta^i,\varrho^{i})
&=&
\lambda_0(t)\exp\bigg(\sum_{j\in\bbJ}
\vartheta^i_jX_j(t)\bigg)\>p^{k_i}_i(t,\varrho^i)
\label{eq:MarkedIntensity}
\eeas
at time $t$ for $(i,k_i)\in\check{\bbI}$, where 
$\vartheta^i=(\vartheta^i_j)_{j\in\bbJ}$ ($i\in\bbI$) and $\varrho^i$ ($i\in\bbI$) are unknown parameters. 
More precisely, given a probability space $(\Omega,\calf,P)$ equipped with 
a right-continuous filtration $\bbF=(\calf_t)_{t\in\bbR_+}$, 
$\lambda_0=(\lambda_0(t))_{t\in\bbR_+}$ is a non-negative predictable process, 
$X_j=(X_j(t))_{t\in\bbR_+}$ is a predictable process for each $j\in\bbJ=\{1,...,\bar{j}\}$, 
and $p_i^{k_i}(t,\rho^i)$ is a non-negative predictable process for each $(i,k_i)\in\check{\bbI}$. 
Later we will put a condition so that the mapping $t\mapsto\lambda^{i,k_i}(t,\vartheta^i,\varrho^i)$ 
is locally integrable with respect to $dt$, 
and we assume that $N^{i,k_i}_0=0$, and for each $(i,k_i)\in\check{\bbI}$, the process 
\beas 
N^{i,k_i}_t-\int_0^t\lambda^{i,k_i}(s,(\vartheta^i)^*,(\varrho^i)^*)ds
\eeas
is a local martingale for a value $\big((\vartheta^i)^*,(\varrho^i)^*\big)$ of 
the parameter $\big(\vartheta^i,\varrho^i\big)$. 

In what follows, we consider the processes $p^{k_i}_i(t,\varrho^i)$ such that 
\bea\label{201912012343}
\sum_{k_i\in\bbK_i}p^{k_i}_i(t,\varrho^i)&=&1 
\eea
for $i\in\bbI$. 
Then the $\bar{k}_i$-dimensional process $(p^{k_i}_i(t,\varrho^i))_{k_i\in\bbK_i}$ gives 
the conditional distribution of the event $k_i$ when the event $i$ occurred. 
Under (\ref{201912012343}), the intensity process of $N^i$ becomes 
\bea\label{201912012346} 
\lambda^i(t,\vartheta^i)
&=&
\sum_{k_i\in\bbK_i}\lambda^{i,k_i}(t,\vartheta^i,\varrho^i)
\yeq 
\lambda_0(t)\exp\bigg(\sum_{j\in\bbJ}\vartheta^i_jX_j(t)\bigg).
\eea

The process $\lambda_0$ is called a baseline intensity, whose structure will not be specified, 
in other words, $\lambda_0$ will be treated as a nuisance parameter, 
differently from the use of Cox regression as in \cite{MuniTokeYoshida2017}. 
The baseline intensity may represent the global market activity in finance, for example, and 
its irregular change may limit the reliability of estimation procedures and predictions for any model fitted to it. 
\cite{MuniTokeYoshida2019} took an approach with an unstructured baseline intensity process and showed 
advantages of such modeling. 
Statistically, the process $\bbX(t)=(X_j(t))_{j\in\bbJ}$ is an observable covariate process. 
Since the effect of these covariate processes to the amplitude of $\lambda^i(t,\vartheta^i)$ 
is contaminated by the unobservable and structurally unknown baseline intensity, 
a more interesting measure of dependency of $\lambda^i(t,\vartheta^i)$ to $\bbX(t)$ is the ratio 
\beas 
\lambda^i(t,\vartheta^i)/\sum_{i'\in\bbI}\lambda^{i'}(t,\vartheta^{i'})
\eeas
for $i\in\bbI$. 
Thus, we introduce the difference parameters $\theta^i_j=\vartheta^i_j-\vartheta^0_j$ ($i\in\bbI,\>j\in\bbJ$), 
($\theta^0_j=0$ in particular) 
and consider the ratios 
\bea\label{201909091443} 
r^i(t,\theta) 
&=& 
\frac{\exp\bigg(\sum_{j\in\bbJ}\vartheta^i_jX_j(t)\bigg)}
{\sum_{i'\in\bbI}\exp\bigg(\sum_{j\in\bbJ}\vartheta^{i'}_jX_j(t)\bigg)}
\yeq
\frac{\exp\bigg(\sum_{j\in\bbJ}\theta^i_jX_j(t)\bigg)}
{1+\sum_{i'\in\bbI_0}\exp\bigg(\sum_{j\in\bbJ}\theta^{i'}_jX_j(t)\bigg)}
\eea
for $i\in\bbI$, 
where $\theta=(\theta^i_j)_{i\in\bbI_0,j\in\bbJ}$ with $\bbI_0=\bbI\setminus\{0\}=\{1,...,\bar{i}\}$. 

In this paper, we further assume that the factor $p^{k_i}_i(t,\varrho^i)$ is given by
\beas
p^{k_i}_i(t,\varrho^i)
&=& 
\frac{\exp\bigg(\sum_{j_i\in\bbJ_i}\varrho^{i,k_i}_{j_i}Y^i_{j_i}(t)\bigg)}
{\sum_{k_i'\in\bbK_i}\exp\bigg(\sum_{j_i\in\bbJ_i}\varrho^{i,k_i'}_{j_i}Y^i_{j_i}(t)\bigg)}
\eeas
for $(i,k_i)\in\check{\bbI}$, $\bbJ_i=\{1,...,\bar{j}_i\}$. 
Obviously, $p^{k_i}_i(t,\varrho^i)=q^{k_i}_i(t,\rho^i)$ defined by 
\bea\label{201909091444} 
q^{k_i}_i(t,\rho^i)
&=& 
\frac{\exp\bigg(\sum_{j_i\in\bbJ_i}\rho^{i,k_i}_{j_i}Y^i_{j_i}(t)\bigg)}
{1+\sum_{k_i'\in\bbK_{i,0}}\exp\bigg(\sum_{j_i\in\bbJ_i}\rho^{i,k_i'}_{j_i}Y^i_{j_i}(t)\bigg)}
\eea
for $(i,k_i)\in\check{\bbI}$, 
where $\rho^{i,k_i}_{j_i}=\varrho^{i,k_i}_{j_i}-\varrho^{i,0}_{j_i}$ $(k_i\in\bbK_i,\>j\in\bbJ_i,\>i\in\bbI)$, 
$\rho^{i,0}_{j_i}=0$ in particular, and 
$\rho^i=(\rho^{i,k_i}_{j_i})_{k_i\in\bbK_{i,0},j_i\in\bbJ_i}$ ($i\in\bbI$) 
with $\bbK_{i,0}=\bbK_i\setminus\{0\}=\{1,...,\bar{k}_i\}$. 
The predictable processes $(Y^i_{j_i}(t))_{t\in\bbR_+}$ ($i\in\bbI,\>j_i\in\bbJ_i$) are observable covariate processes, 
$\bbJ_i$ being a finite index set. 
This is a multinomial logistic regression model. 

Let $\Theta$ be a bounded open convex set in $\bbR^\sfp$ with $\sfp=\bar{i}\>\bar{j}$. 
For each $i\in\bbI$, $\calr_i$ denotes a bounded open convex set in $\bbR^{\sfp_i}$ with $\sfp_i=\bar{j}_i\>\bar{k}_i$. 
Write $\rho=(\rho^i)_{i\in\bbI}$. 
Let $\calr=\Pi_{i\in\bbI}\calr_i$. 
We will consider $\overline{\Theta}\times\overline{\calr}$ as the parameter space of $(\theta,\rho)$. 

\begin{remark}
The marked ratio model 
\beas 
\lambda^{i,k_i}(t,\vartheta^i,\varrho^{i})
&=&
\lambda_0(t)\exp\bigg(\sum_{j\in\bbJ}
\vartheta^i_jX_j(t)\bigg)\>
\frac{\exp\bigg(\sum_{j_i\in\bbJ_i}\varrho^{i,k_i}_{j_i}Y^i_{j_i}(t)\bigg)}
{\sum_{k_i'\in\bbK_i}\exp\bigg(\sum_{j_i\in\bbJ_i}\varrho^{i,k_i'}_{j_i}Y^i_{j_i}(t)\bigg)}
\eeas
is in general not equivalent to a non-marked ratio model in larger dimension, in which we would write the intensity of the counting process of events of type $i\in\mathbb I$ with mark $k_i\in\mathbb K_i$ as
\beas 
\lambda^{i,k_i}(t,\vartheta^{i,k_i})
&=&
\tilde\lambda_0(t)\exp\bigg(\sum_{j\in\tilde\bbJ}
\vartheta^{i,k_i}_jZ_j(t)\bigg).
\eeas
for some covariate processes $Z_j, j\in\tilde\bbJ$.
Equivalence of the models would require these expressions to coincide for some sets of covariates and parameters. However, if $Z_j(t)=0$ for all $j\in\tilde\bbJ$, then necessarily $X_j(t)=0$ for all $j\in\bbJ$ and $Y^i_{j_i}(t)=0$ for all $i\in\mathbb I$ and $j_i\in\mathbb J_i$. This in turn implies $ \frac{1}{|\mathbb K_i|} = \frac{\tilde\lambda_0(t)}{\lambda_0(t)}$ for all $i\in\mathbb I$, which is generally not true.
In Section \ref{subsec:Prediction}, a non-marked ratio model is used as a benchmark to assess the performances of the marked ratio model. Prediction performances are indeed shown to be different.
\end{remark}

\section{Quasi-likelihood estimation of two-step ratio model}
\label{sec:QuasiLikelihoodEstimation}

\subsection{Quasi-maximum likelihood estimator and quasi-Bayesian estimator}
\label{subsec:QuasiLikelihoodEstimators}
The two step marked ratio model consists of the two kinds of ratio models 
(\ref{201909091443}) and (\ref{201909091444}). 
Estimation of this model can be carried out with multiple successive ratio models. 

In the first step, we consider the parameter $\theta=(\theta^i_j)_{i\in\bbI_0,j\in\bbJ}$ and 
the ratios (\ref{201909091443}) for $i\in\bbI$. 
The quasi-log-likelihood based on observations on $[0,T]$ for this ratio model is
\bea
\bbH_T(\theta) &=& \sum_{i\in\bbI}\int_0^T\log r^i(t,\theta)\>dN^i_t. 
\label{eq:ThetaQL}
\eea
A quasi-maximum likelihood estimator (QMLE) for $\theta$ is a measurable mapping 
$\hat{\theta}_T^M:\Omega\to\overline{\Theta}$ satisfying 
\beas 
\bbH_T(\hat{\theta}_T^M) &=& \max_{\theta\in\overline{\Theta}}\bbH_T(\theta)
\eeas
for all $\omega\in\Omega$. 
\footnote{Originally, $\hat{\theta}_T^M$ is defined on a sample space ${\sf S}_T$ expressing all the possible outcomes   
of $(\lambda_0(t),X_j(t),Y^i_{j_i}(t);\>t\in[0,T],i\in\bbI, j\in\calj, j_i\in\bbJ_i)$. 
If $(\Omega,\calf,P)$ is an abstract space used for defining the true probability measure $P^*_T$ on ${\sf S}_T$ 
by some random variable $V_T:\Omega\to{\sf S}_T$ (i.e. $P^*_T=PV_T^{-1}$), then 
treating $\hat{\theta}_T^M$ as a function on $\Omega$ conflicts with the definition of $\hat{\theta}_T^M$. 
However, what we want to investigate is concerning the distribution of $\hat{\theta}_T^M$ (defined on $S_T$) 
under $P^*_T$, and then we can pull back $\hat{\theta}_T^M$ on $S_T$ to $\Omega$ by $V_T$ 
if $P^*_T=PV_T^{-1}$. For this reason, we can identify $\hat{\theta}_T^M$ with $\hat{\theta}_T^M\circ V_T$, 
and may regard $\hat{\theta}_T^M$ as defined on $\Omega$. 
This remark makes sense especially when one treats a weak solution of a stochastic differential equation 
for a covariate. 
}

In the second step, we consider the ratios (\ref{201909091444}) and the associated quasi-log-likelihood
\bea
\bbH_T^{(i)}(\rho^i)
&=& 
\sum_{k_i\in\bbK_i}
\int_0^T\log q^{k_i}_i(t,\rho^i)\>dN^{i,k_i}_t.
\label{eq:RhoQL}
\eea
for $i\in\bbI$. 
Then a measurable mapping $\hat{\rho}^{i,M}_T:\Omega\to\overline{\calr}_i$ is called 
a quasi-maximum likelihood estimator (QMLE) for $\rho^i$ if 
\beas 
\bbH_T^{(i)}(\hat{\rho}^{i,M}_T) &=& \max_{\rho^i\in\overline{\calr}_i}\bbH_T^{(i)}(\rho^i). 
\eeas

It is possible to pool these estimating functions by the single estimating function 
\bea\label{201909120239} 
\bbH_T(\theta,\rho)
&=& 
\bbH_T(\theta)
+
\sum_{i\in\bbI}\bbH_T^{(i)}(\rho^i).
\eea
In other words, 
\bea\label{201911301554}
\bbH_T(\theta,\rho) 
&=& 
\sum_{i\in\bbI}\sum_{k_i\in\bbK_i}
\int_0^T\log \big(r^i(t,\theta)q^{k_i}_i(t,\rho^i)\big)\>dN^{i,k_i}_t
\eea
The collection of QMLEs $\big(\hat{\theta}_T^B,(\hat{\rho}_T^{i,M})_{i\in\bbI}\big)$ 
is a QMLE for $\bbH_T(\theta,\rho)$. 
Use of $\bbH_T(\theta,\rho)$ is convenient when we consider asymptotic distribution of 
the estimators $\hat{\theta}_T^M$ and $\hat{\rho}_T^i$ ($i\in\bbI$) jointly. 

The quasi-Bayesian estimator (QBE) $\big(\hat{\theta}_T^B,(\hat{\rho}_T^{i,B})_{i\in\bbI}\big)$ is defined by 
\bea\label{201912020251} 
\hat{\theta}_T^B
&=&
\bigg[\int_{\Theta\times\calr}\exp\big(\bbH_T(\theta,\rho)\big)\>\varpi(\theta,\rho)\>d\theta d\rho\bigg]^{-1}
\int_{\Theta\times\calr}\theta\exp\big(\bbH_T(\theta,\rho)\big)\>\varpi(\theta,\rho)\>d\theta d\rho
\eea
and 
\bea\label{201912020252} 
\hat{\rho}^{i,B}_T
&=&
\bigg[\int_{\Theta\times\calr}\exp\big(\bbH_T(\theta,\rho)\big)\>\varpi(\theta,\rho)\>d\theta d\rho\bigg]^{-1}
\int_{\Theta\times\calr}\rho^i\exp\big(\bbH_T(\theta,\rho)\big)\>\varpi(\theta,\rho)\>d\theta d\rho
\eea
for a prior probability density $\varpi(\theta,\rho)$ on $\Theta\times\calr$. 
We assume that $\varpi:\Theta\times\calr\to\bbR_+$ is continuous and 
\bea\label{201912020253} 
0<\inf_{(\theta,\rho)\in\Theta\times\calr}\varpi(\theta,\rho)
\leq\sup_{(\theta,\rho)\in\Theta\times\calr}\varpi(\theta,\rho)<\infty. 
\eea

Since $\bbH_T(\theta)$ and $\bbH_T^{(i)}(\rho^i)$ have no common parameters, 
the maximization of $\bbH_T(\theta,\rho)$ with respect to 
the parameters $\theta$ and $\rho^i$ $(i\in\bbI)$ can be carried out separately. 
However, these components are not always individually treated 
for the QBE. 
If $\varpi(\theta,\rho)$ is a product of prior densities as 
$\varpi(\theta,\rho)=\varpi'(\theta)\Pi_{i\in\bbI}\varphi^i(\rho^i)$, then 
the each integral in (\ref{201912020251}) and (\ref{201912020252}) is simplified and 
we can compute $\hat{\theta}_T^B$ and $\hat{\rho}_T^{i,B}$ ($i\in\bbI$) separately: 
\beas 
\hat{\theta}_T^B
&=&
\bigg[\int_{\Theta}\exp\big(\bbH_T(\theta)\big)\>\varpi'(\theta)\>d\theta\bigg]^{-1}
\int_{\Theta}\theta\exp\big(\bbH_T(\theta)\big)\>\varpi'(\theta)\>d\theta
\eeas
and 
\beas
\hat{\rho}^{i,B}_T
&=&
\bigg[\int_{\calr_i}\exp\big(\bbH_T^{(i)}(\rho^i)\big)\>\varpi^i(\rho^i)\>d\rho^i\bigg]^{-1}
\int_{\calr_i}\rho^i\exp\big(\bbH_T^{(i)}(\rho^i)\big)\>\varpi^i(\rho^i)\>d\rho^i
\eeas
for $i\in\bbI$. 

\subsection{Quasi-likelihood analysis}
\label{subsec:QuasiLikelihoodAnalysis}
\begin{en-text}
\bi
\im Assumption. There exists a positive constant $\chi_0$ such that 
\beas 
\bbY(\theta,\rho)
&\leq& -\chi_0\big|(\theta,\rho)-(\theta^*,\rho^*)\big|^2
\eeas
for all $(\theta,\rho)$. 
{\bf But this condition follows from a local condition?!}
\ei
\end{en-text}
Let $\bbX(t)=(X_j(t))_{j\in\bbJ}$ and let 
$\bbY^i(t)=\big(Y^i_{j_i}(t)\big)_{j_i\in\bbJ_i}$ for $i\in\bbI$.
%
We consider the following conditions. 
\begin{description}
\item[{[M1]}] The process $\big(\lambda_0(t), \bbX(t),\bbY(t))$ is a stationary process and the random variables $\lambda_0(0)$, $\exp(|X_j(0)|)$ and $\exp(|Y^i_{j_i}(0)|)$ are in $L^\inftym=\cap_{p>1}L^p$ for $j\in\bbJ$, $j_i\in\bbJ_i$ and $i\in\bbI$. 
\end{description}

The alpha mixing coefficient $\alpha(h)$ is defined by 
\beas 
\alpha(h)
&=&
\sup_{t\in\bbR_+}\sup_{\genfrac{}{}{0pt}{}{A\in\calb_{[0,t]}}{B\in\calb_{[t+h,\infty)}}}
\big|P[A\cap B]-P[A]P[B]\big|,
\eeas
where for $I\subset\bbR_+$, $\calb_I$ denotes the $\sigma$-field 
generated by $\big(\lambda_0(t),(X_j(t))_{j\in\bbJ},(Y^{i,k_i}_{j_i}(t))_{i\in\bbI,j_i\in\bbJ_i,k_i\in\bbK_i}\big)$. 

\bd\item[{[M2]}]
The alpha mixing coefficient $\alpha(h)$ is rapidly decreasing 
in that $\alpha(h)h^L\to0$ as $h\to\infty$ for every $L>0$. 
\ed
\halflineskip

In the two-step ratios model, the category $(i,k_i)$ is selected 
with two-fold multinomial distributions of sample size equal to $1$. 
First the class $i\in\bbI$ is selected when $\xi_i=1$ for some random variable 
\beas 
\xi=(\xi_0,...,\xi_{\bar{i}})\sim\text{Multinomial}(1;\pi_0,....,\pi_{\bar{i}}). 
\eeas
If $\xi_i=1$ for a class $i\in\bbI$, then the class $k_i\in\bbK_i$ is chosen as $k_i=k$ 
when $\eta^i_k=1$ for some independent random variable 
\beas
\eta^i=(\eta^i_0,...,\eta^i_{\bar{k}_i})\sim\text{Multinomial}(1;\pi_0',...,\pi'_{\bar{k}_i}). 
\eeas

Denote by ${\sf V}(x,\theta)$ 
the variance matrix of the $(1+\overline{i})$-dimensional multinomial distribution 
$\text{M}(1;\pi_0,\pi_1,...,\pi_{\overline{i}})$ with 
$\pi_i=\dot{r}^i(x,\theta)$, $i\in\mathbb I$, 
where 
\beas 
\dot{r}^i(x,\theta)
&=&
\frac{\exp\bigg(\sum_{j\in\bbJ}\vartheta^i_jx_j\bigg)}
{\sum_{i'\in\bbI}\exp\bigg(\sum_{j\in\bbJ}\vartheta^{i'}_jx_j\bigg)}
\yeq
\frac{\exp\bigg(\sum_{j\in\bbJ}\theta^i_jx_j\bigg)}
{1+\sum_{i'\in\bbI_0}\exp\bigg(\sum_{j\in\bbJ}\theta^{i'}_jx_j\bigg)},
\quad x=(x_j)_{j\in\bbJ}
\eeas
Denote by ${\sf V}^i(x,\rho^i)$ 
the variance matrix of the $(1+\overline{k}_i)$-dimensional multinomial distribution 
$\text{M}(1;\pi_0',\pi_1',...,\pi_{\overline{k}_i}')$ with 
$\pi_{k_i}'=\dot{q}_i^{k_i}(y^i,\rho^i)$, $k_i\in\bbK_i$, 
where
\begin{align*}
\dot{q}_i^{k_i}(y^i,\rho^i)
& =
\frac{\exp\bigg(\sum_{j_i\in\bbJ_i}\varrho^{i,k_i}_{j_i}y^i_{j_i}\bigg)}
{\sum_{k_i'\in\bbK_i}\exp\bigg(\sum_{j_i\in\bbJ_i}\varrho^{i,k_i'}_{j_i}y^i_{j_i}\bigg)}
\\
& =
\frac{\exp\bigg(\sum_{j_i\in\bbJ_i}\rho^{i,k_i}_{j_i}y^i_{j_i}\bigg)}
{1+\sum_{k_i'\in\bbK_{i,0}}\exp\bigg(\sum_{j_i\in\bbJ_i}\rho^{i,k_i'}_{j_i}y^i_{j_i}\bigg)},
\quad y^i=(y^i_{j_i})\in\bbR^{\bar{j}_i}\quad(i\in\bbI).
\end{align*}

Let us introduce some notations used in the following analysis. For a tensor ${\sf T}=({\sf T}_{i_1,...,i_k})_{i_1,...,i_k}$, we write 
\begin{equation}
{\sf T}[u_1,...,u_k]
=
{\sf T}[u_1\otimes\cdots\otimes u_k]
=
\sum_{i_1,...,i_k}{\sf T}_{i_1,...,i_k}
u_1^{i_1}\cdots u_k^{i_k}
\end{equation}
for $u_1=(u_1^{i_1})_{i_1}$,..., $u_k=(u_k^{i_k})_{i_k}$. Brackets $[\ ,..., \ ]$ stand for a multilinear mapping. 
We denote by $u^{\otimes r}=u\otimes\cdots\otimes u$ the $r$ times tensor product of $u$.

Let 
\beas 
\Gamma_T(\theta,\rho) &=& -T^{-1}\partial_{(\theta,\rho)}^2\bbH_T(\theta,\rho)
\eeas
and let $\Gamma_T=\Gamma_T(\theta^*,\rho^*)$. 
Then, as detailed on p. \pageref{Gamma}, 
\beas 
\Gamma_T(\theta,\rho)
&=& 
\text{diag}\big[
\Gamma_T(\theta),\Gamma_T^1(\rho^1),...,\Gamma_T^{\bar{i}}(\rho^{\bar{i}})
\big]
\eeas
where 
\bea\label{171022-1} 
\Gamma_T(\theta) [u^{\otimes2}]
= 
\frac{1}{T}\int_0^T \bigg({\sf V}_0(\mathbb X(t),\theta)\otimes\mathbb X(t)^{\otimes2}\bigg)
[u^{\otimes2}]\sum_{i\in\mathbb I}dN^i_t
\quad(u\in\bbR^{\sfp})
\eea
with ${\sf V}_0(x,\theta)=({\sf V}(x,\theta)_{i,i'})_{i,i'\in\mathbb I_0}$,
and 
\beas 
\Gamma^i_T(\rho^i) [(u^i)^{\otimes2}]
= 
\frac{1}{T}\int_0^T \bigg({\sf V}^i_0(\bbY^i(t),\rho^i)\otimes\bbY^i(t)^{\otimes2}\bigg)
[(u^i)^{\otimes2}]dN^i_t
\quad(u^i\in\bbR^{\sfp_i})
\eeas
with ${\sf V}^i_0(y^i,\rho^i)=({\sf V}^i(y^i,\rho^i)_{k_i,k_i'})_{k_i,k_i'\in\bbK_{i,0}}$.

Let 
\begin{equation} 
\Lambda(w,x) =
w\sum_{i\in\mathbb I}\exp\big(x\big[\vartheta^{*i}\big]\big)
\end{equation}
for $w\in\mathbb R_+$ and $x\in\mathbb R^{\overline{j}}$. 

\begin{en-text}
\im 
For $i\in\bbI$, let \koko
\begin{equation} 
\Lambda^i(w,{\colorr y^i}) =
w\exp\big({\colorr y^i}\big[\vartheta^{*i}\big]\big)
\end{equation}
for $w\in\mathbb R_+$ and ${\colorr y^i\in\mathbb R^{\overline{k}_i}}$. 
\end{en-text}

We have
\beas
{\sf V}(x,\theta)_{i,i'}
&=&
1_{\{i=i'\}}\dot{r}^i(x,\theta)-\dot{r}^i(x,\theta)\dot{r}^{i'}(x,\theta)
\eeas
Therefore
\bea\label{201912010320}
{\sf V}(\bbX(t),\theta)_{i,i'}
&=&
1_{\{i=i'\}}r^i(t,\theta)-r^i(t,\theta)r^{i'}(t,\theta)
\eea
and ${\sf V}_0(\bbX(t),\theta)_{i,i'}={\sf V}(\bbX(t),\theta)_{i,i'}$ for $i,i'\in\bbI_0$.
Write
${\sf V}_0(x)={\sf V}_0(x,\theta^*)$. 

We have 
\beas
{\sf V}^i(y^i,\rho^i)_{k_i,k_i'} 
&=& 
1_{\{k_i=k_i'\}}\dot{q}^{k_i}(y^i,\rho^i)-\dot{q}^{k_i}(y^i,\rho^i)\dot{q}^{k_i'}(y^i,\rho^i).
\eeas
Hence
\bea\label{201912010321} 
{\sf V}^i(\bbY^i(t),\rho^i)_{k_i,k_i'} 
&=&
1_{\{k_i=k_i'\}}q^{k_i}_i(t,\rho^i)-q^{k_i}_i(t,\rho^i)q^{k_i'}_i(t,\rho^i)
\eea
and ${\sf V}_0^i(\bbY^i(t),\rho^i)_{k_i,k_i'}={\sf V}^i(\bbY^i(t),\rho^i)_{k_i,k_i'}$ 
for $k_i,k_i'\in\bbK_{i,0}$. 
We denote  
${\sf V}^i_0(y^i)={\sf V}^i_0(y^i,(\rho^i)^*)$. 

Let 
\beas 
\Gamma(\theta)[u^{\otimes2}]
&=&
E\bigg[\bigg({\sf V}_0(\mathbb X(0),\theta)\otimes \mathbb X(0)^{\otimes2}\bigg)[u^{\otimes2}]
\Lambda(\lambda_0(0),\mathbb X(0))\bigg]
\eeas
for $u\in\bbR^\sfp$,  
and let 
\beas 
\Gamma^i(\rho^i)[(u^i)^{\otimes2}]
&=&
E\bigg[\bigg({\sf V}^i_0(\bbY^i(0),\rho^i)\otimes \bbY^i(0)^{\otimes2}\bigg)[(u^i)^{\otimes2}]
 \Lambda(\lambda_0(0),\bbX(0))r^i(0,\theta^*)\bigg]
\eeas
for $u^i\in\bbR^{\sfp_i}$, $i\in\bbI$.  
Let $\check{\sfp}=\sfp+\sum_{i\in\bbI}\sfp_i=\bar{i}\>\bar{j}+\sum_{i\in\bbI}\bar{k}_i\bar{j}_i$. 
The full information matrix is the $\check{\sfp}\times\check{\sfp}$ block diagonal matrix is 
\beas 
\Gamma(\theta,\rho)
&=&
\text{diag}\big[\Gamma(\theta),\Gamma^0(\rho^0),
\Gamma^1(\rho^1),...,\Gamma^{\bar{i}}(\rho^{\bar{i}})\big],
\eeas
and in particular set 
\begin{equation}
\Gamma =  \Gamma(\theta^*,\rho^*).
\label{eq:MatrixGamma}
\end{equation}
\noindent

An identifiability condition will be imposed. \halflineskip
\bd\item[{[M3]}]
$\ds \inf_{\theta\in\Theta}\inf_{u\in\bbR^\sfp:\>|u|=1}\Gamma(\theta)[u^{\otimes2}]>0$ 
and 
$\ds \inf_{\rho^i\in\calr_i}\inf_{u\in\bbR^{\sfp_i}:\>|u^i|=1}\Gamma^i(\rho^i)[(u^i)^{\otimes2}]>0$ 
for every $i\in\bbI$. 
\ed
\halflineskip

For the QMLE $\hat{\psi}_T^M=(\hat{\theta}^M_T,\hat{\rho}^M_T)$ and the 
QBE $\hat{\psi}_T^B=(\hat{\theta}^B_T,\hat{\rho}^B_T)$ of 
$\psi=(\theta,\rho)=(\theta,\rho^1,...,\rho^{\bar{i}})$, 
let 
\beas 
\hat{u}_T^{\sf A}&=& T^{1/2}\big(\hat{\psi}^{\sf A}-\psi^*)
\qquad({\sf A}\in\{M,B\}). 
\eeas
\begin{theorem}\label{201911220134}
Suppose that Conditions $[M1]$, $[M2]$ and $[M3]$ are satisfied. 
Then 
\beas 
E[f(\hat{u}_T^{\sf A})]&\to& \bbE[f(\Gamma^{-1/2}\zeta)]
\eeas
as $T\to\infty$ for ${\sf A}\in\{M,B\}$ and every $f\in C(\bbR^{\check{\sfp}})$ 
at most polynomial growth, 
where $\zeta$ is a $\check{\sfp}$-dimensional standard Gaussian random vector.
\label{thm:ConvergenceResult} 
\end{theorem}

\begin{example}\label{ex:NumericalApplication}
As an illustration we consider the case with two processes ($\mathbb I=\{0,1\}$), and two marks for each process ($\mathbb K_0=\mathbb K_1=\{0,1\}$). The first state-dependent term takes into account one covariate $X_1$ (i.e. $\mathbb J=\{1\}$). The mark distributions both depend on another covariate $Y_1$ (i.e. $\mathbb J_0=\mathbb J_1=\{1\}$).
In this example, we assume that $X_1$ and $Y_1$ are independent Markov chains with values in $\{-1,1\}$ and constant transition intensities $\lambda_X$ and $\lambda_Y$. We assume that $\lambda_0$ is the intensity of a Hawkes process $(H_t)_{t\geq 0}$ with a single exponential kernel, i.e. $\lambda_0(t)=\mu+\int_{0}^t \alpha e^{-\beta (t-s)\,dH_s}$, with $(\alpha,\beta)\in(\mathbb R_ +^*)^2, \frac{\alpha}{\beta}<1$.

The two-step ratio model estimates the parameters $(\theta^1_1,\rho^{0,1}_1,\rho^{1,1}_1)$ defined as $\theta^1_1=\vartheta^1_1-\vartheta^0_1$ and $\rho^{i,1}_1=\varrho^{i,1}_1-\varrho^{i,0}_1$, $i=0,1$.
In this specific case the matrix $\Gamma$ of Equation \eqref{eq:MatrixGamma} is a $3\times 3$-diagonal matrix, and a direct computation shows that the diagonal coefficients are
\begin{align*}
	\Gamma_{0,0} & = \frac{\mu}{1-\frac{\alpha}{\beta}} \frac{e^{\theta^1_1}}{1+e^{\theta^1_1}}\left(\cosh \vartheta^0_1 + \cosh \vartheta^1_1 \right),
	\\
	\Gamma_{1,1} & = \frac{\mu}{1-\frac{\alpha}{\beta}} \frac{e^{\rho^{0,1}_1}}{1+e^{\rho^{0,1}_1}} \frac{e^{\theta^1_1/2}}{1+e^{\theta^1_1}}\left(\cosh\frac{\vartheta^0_1+\vartheta^1_1}{2}+\cosh\frac{3\vartheta^1_1-\vartheta^0_1}{2}\right),
	\\
	\Gamma_{2,2} & = \frac{\mu}{1-\frac{\alpha}{\beta}} \frac{e^{\rho^{1,1}_1}}{1+e^{\rho^{1,1}_1}} \frac{e^{\theta^1_1/2}}{1+e^{\theta^1_1}}\left(\cosh\frac{\vartheta^0_1+\vartheta^1_1}{2}+\cosh\frac{3\vartheta^1_1-\vartheta^0_1}{2}\right).
\end{align*}

We run 1000 simulations of the processes $(N^0, N^1)$ with their marks for various values of horizon $T$. Numerical values used in these simulations are the following: $\mu=0.5$, $\alpha=1.0$, $\beta = 2.0$, $\lambda_X=\lambda_Y=0.5$, $\vartheta^0_1=-0.75$, $\vartheta^1_1=0.75$, $\varrho^{0,0}_1=-0.5$, $\varrho^{0,1}_1=0.5$, $\varrho^{1,0}_1=-1.0$, $\varrho^{1,1}_1=1.0$.
For each simulation, we compute the quasi-maximum likelihood estimators $(\hat\theta^1_1,\hat\rho^{0,1}_1,\hat\rho^{1,1}_1)$ with the two-step ratios described above. Table \ref{table:NumericalApplication} gives the mean estimators and the true values of the parameters, as well as the empirical standard deviation, compared to the theoretical values $T^{-\frac{1}{2}}\Gamma_{i,i}^{-\frac{1}{2}}$, $i=0,1,2$ from Theorem \ref{thm:ConvergenceResult}, for various values of $T$.
\begin{table}[htbp]
\centering
\begin{tabular}{|c|c|c|c|c|}
\hline
\multicolumn{ 1}{|c|}{}  &  & $\theta^1_1$ & $\rho^{0,1}_1$ & $\rho^{1,1}_1$
\\ 
\multicolumn{ 1}{|c|}{T}  & \textbf{True Value} & \textbf{1.500} & \textbf{1.000} & \textbf{2.000} 
\\ \hline
\multicolumn{ 1}{|c|}{} & Estimator mean & 1.817 & 1.576 & 5.146
\\ 
\multicolumn{ 1}{|c|}{10} & Estimator sd & \textit{1.829} & \textit{2.875} & \textit{5.491} 
\\ 
\multicolumn{ 1}{|c|}{} & $T^{-\frac{1}{2}}\Gamma_{i,i}^{-\frac{1}{2}}$ & \textit{0.509} & \textit{0.627} & \textit{0.858}
\\ \hline
\multicolumn{ 1}{|c|}{} & Estimator Mean & 1.541 & 1.044 & 2.402
\\ 
\multicolumn{ 1}{|c|}{30} & Estimator sd & \textit{0.324} & \textit{0.610} & \textit{2.096}
\\ 
\multicolumn{ 1}{|c|}{} & $T^{-\frac{1}{2}}\Gamma_{i,i}^{-\frac{1}{2}}$ & \textit{0.294} & \textit{0.362} & \textit{0.495}
\\ \hline
\multicolumn{ 1}{|c|}{} & Estimator Mean & 1.508 & 0.999 & 2.011
\\ 
\multicolumn{ 1}{|c|}{100} & Estimator sd & \textit{0.164} & \textit{0.201} & \textit{0.289}
\\ 
\multicolumn{ 1}{|c|}{} & $T^{-\frac{1}{2}}\Gamma_{i,i}^{-\frac{1}{2}}$ & \textit{0.161} & \textit{0.198} & \textit{0.271} 
\\ \hline
\multicolumn{ 1}{|c|}{} & Estimator Mean & 1.502 & 1.005 & 2.013
\\ 
\multicolumn{ 1}{|c|}{300} & Estimator sd & \textit{0.094} & \textit{0.114} & \textit{0.159}
\\ 
\multicolumn{ 1}{|c|}{} & $T^{-\frac{1}{2}}\Gamma_{i,i}^{-\frac{1}{2}}$ & \textit{0.093} & \textit{0.114} & \textit{0.157}
\\ \hline
\multicolumn{ 1}{|c|}{} & Estimator Mean & 1.501 & 1.000 & 2.009
\\ 
\multicolumn{ 1}{|c|}{1000} & Estimator sd & \textit{0.052} & \textit{0.065} & \textit{0.085}
\\ 
\multicolumn{ 1}{|c|}{} & $T^{-\frac{1}{2}}\Gamma_{i,i}^{-\frac{1}{2}}$ & \textit{0.051} & \textit{0.063} & \textit{0.086}
\\ \hline
\multicolumn{ 1}{|c|}{} & Estimator Mean & 1.498 & 1.001 & 1.999
\\ 
\multicolumn{ 1}{|c|}{3000} & Estimator sd & \textit{0.029} & \textit{0.038} & \textit{0.053}
\\ 
\multicolumn{ 1}{|c|}{} & $T^{-\frac{1}{2}}\Gamma_{i,i}^{-\frac{1}{2}}$ & \textit{0.029} & \textit{0.036} & \textit{0.050}
\\ \hline
\end{tabular}
\caption{Numerical results for the estimation of the model of Example \ref{ex:NumericalApplication}.}
\label{table:NumericalApplication}
\end{table}
For completeness, Figure \ref{fig:NumericalApplication} also plots the empirical standard deviations of the three estimators and the theoretical standard deviation $T^{-\frac{1}{2}}\Gamma_{i,i}^{-\frac{1}{2}}$, $i=0,1,2$ of Theorem \ref{thm:ConvergenceResult}, as a function of the horizon $T$.
\begin{figure}
\centering
\begin{tabular}{ccc}
\includegraphics[width=0.3\textwidth, page=1]{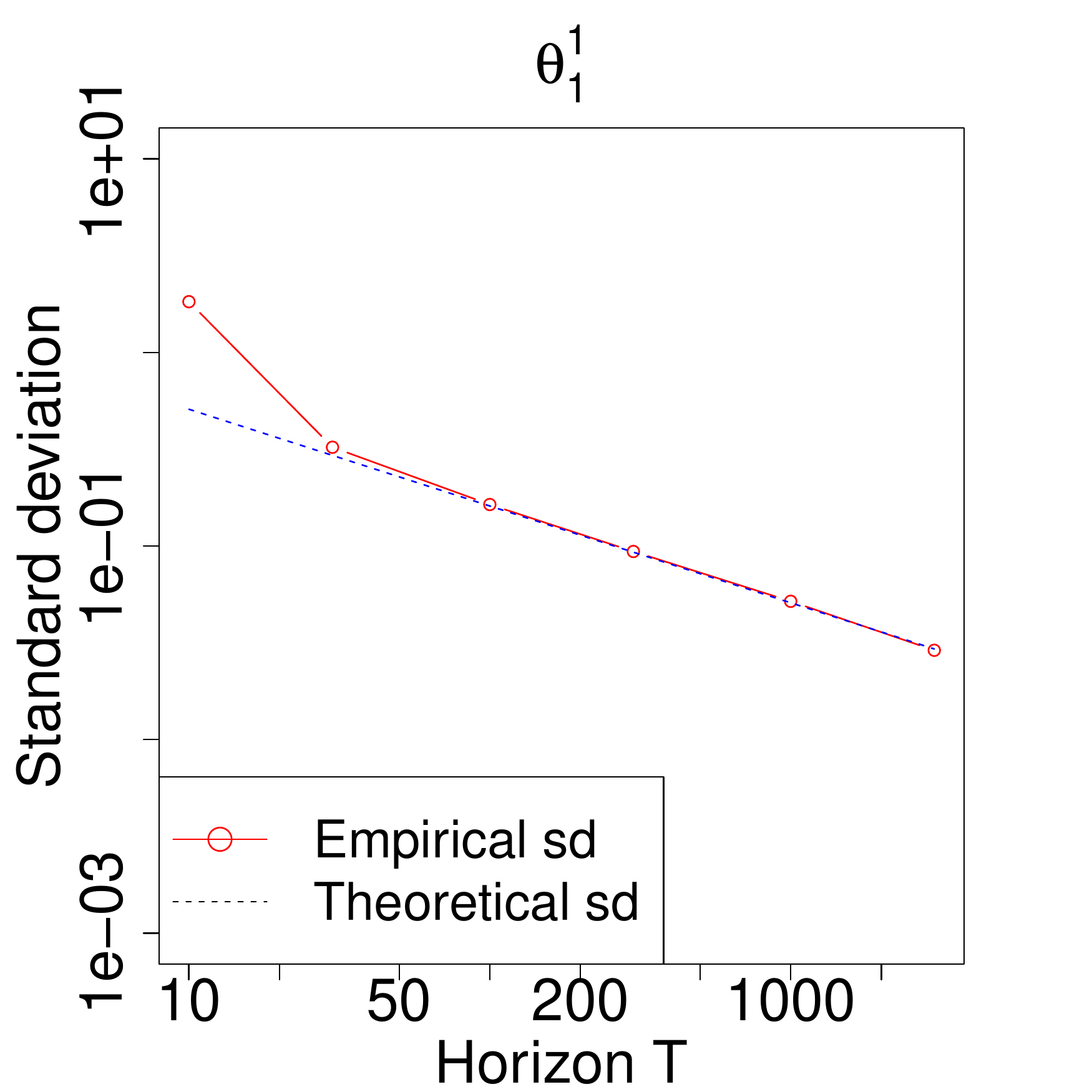}
&
\includegraphics[width=0.3\textwidth, page=2]{NumericalApplication.pdf}
&
\includegraphics[width=0.3\textwidth, page=3]{NumericalApplication.pdf}
\end{tabular}
\caption{Empirical and theoretical standard deviation of the quasi-maximum likelihood estimators $\hat\theta^1_1$ (left), $\hat\rho^{0,1}_1$ (center) and $\hat\rho^{1,1}_1)$ (right).}
\label{fig:NumericalApplication}
\end{figure}
Asymptotic values predicted by Theorem \ref{thm:ConvergenceResult} are indeed empirically retrieved, which ends this numerical illustration.
\end{example}

\section{Modeling and predicting sign and aggressiveness of market orders}
\label{sec:TradeAggressiveness}

\subsection{Intensities of the processes counting market orders}

We consider the market orders submitted to a given limit order book. Let $N^0$ be the process counting the market orders submitted on the bid side (sell market orders) and $N^1$ the process counting the market orders submitted on the ask side (buy market orders). On each side, we further consider whether the order is an aggressive order that moves the price (labeled with mark $1$), or a non-aggressive order that does not move the price (labeled with mark $0$).

We assume that the intensity of an order of type $i\in\mathbb I=\{0,1\}$ with mark $k_i\in\mathbb K=\mathbb K_0=\mathbb K_1=\{0,1\}$ is
\begin{equation}
	\lambda^{i, k_i}(t, \vartheta^i, \varrho^{i}) = \lambda_0(t) \exp\left( \sum_{j\in\mathbb J} \vartheta^i_j X_j(t) \right) \frac{\exp\left(\sum_{j\in\mathbb J_i} \varrho^{i,k_i}_{j_i} Y^i_j(t) \right)}{\sum_{k'_i\in\mathbb K_i} \exp\left(\sum_{j\in\mathbb J_i} \varrho^{i,k'_i}_j Y^i_j(t) \right)}.
\label{eq:TradesModel}
\end{equation}

In the following applications, we will consider several possible models defined with various sets of covariates $X_j$, $j\in\mathbb J$ and $Y^i_j$, $j\in\mathbb J_i$, $i=0,1$. The tested sets of covariates $X_j$, $j\in\mathbb J$ and $Y^i_j$, $j\in\mathbb J_i$, $i=0,1$ will all be subsets of the following list of possible covariates (besides $Z_0=1$ common to all models):
\begin{itemize}
	\item $Z_1(t) = \frac{q^B(t)-q^A(t)}{q^B(t)-q^A(t)}$ where $q^B(t)$ (resp. $q^A(t)$) is the quantity available at the best bid (resp.ask) at time $t$ (i.e. the imbalance);
	\item $Z_2(t) = \epsilon(t)$, where $\epsilon(t)$ is the sign of the last market order at time $t$ ($1$ for an ask market order, $-1$ for a bid market order ;
	\item $Z_3(t) = \sigma(t)\epsilon(t)$, where $\sigma(t)$ is equal to $1$ if the spread at time $t$ is large (larger than a reference value, taken here to be the median spread of the stock), $-1$ otherwise ; 
	\item $Z_4$: $H^{0,1}(t) = \log\left( \mu^{0,1} + \int_0^t \alpha^{0,1} e^{-\beta^{0,1}(t-s)} dN^{0,1}_s \right)$ (Hawkes covariate for aggressive bid market orders)
	\item $Z_5$: $H^{0,0}(t) = \log\left( \mu^{0,0} + \int_0^t \alpha^{0,0} e^{-\beta^{0,0}(t-s)} dN^{0,0}_s \right)$ (Hawkes covariate for non-aggressive bid market orders)
	\item $Z_6$: $H^{1,1}(t) = \log\left( \mu^{1,1} + \int_0^t \alpha^{1,1} e^{-\beta^{1,1}(t-s)} dN^{1,1}_s \right)$ (Hawkes covariate for aggressive ask market orders)
	\item $Z_7$: $H^{1,0}(t) = \log\left( \mu^{1,0} + \int_0^t \alpha^{1,0} e^{-\beta^{1,0}(t-s)} dN^{1,0}_s \right)$ (Hawkes covariate for non-aggressive ask market orders)
	\item $Z_8$: $H^{0}(t) = \log\left( \mu^{0} + \int_0^t \alpha^{0} e^{-\beta^{0}(t-s)} dN^{0}_s \right)$ (Hawkes covariate for bid market orders)
	\item $Z_9$: $H^{1}(t) = \log\left( \mu^{1} + \int_0^t \alpha^{1} e^{-\beta^{1}(t-s)} dN^{1}_s \right)$ (Hawkes covariate for ask market orders)
\end{itemize}
With these Hawkes covariates, the ratio model can actually be seen as a kind of non-linear Hawkes process.

\subsection{Limit order book data}

We use tick-by-tick data for 36 stocks traded on Euronext Paris. The sample spans the whole year 2015, i.e. roughly 200 trading days for each stock, although some days are missing for some stocks. Table \ref{table:ricList} in Appendix \ref{sec:ricList} lists the stocks investigated and the number of trading days available. Rough data consists in a TRTH (Thomson-Reuters Tick History) databases: for each trading day and each stock, one file lists the transactions (quantities and prices) and one file lists the modifications of the limit order book (level, price and quantities). Timestamps are given with a millisecond precision. Synchronization of both files and reconstruction of the limit order book is carried out with the procedure described in \cite{MuniToke2017}. One strong advantage of the ratio model is that it does not require precise timestamps in itself, since timestamps do not appear explicitly in the quasi-likelihood of the ratios, while fitting other intensity-based models (e.g. Hawkes processes) requires unique precise timestamps for log-likelihood computation. Here, if Hawkes fits are used as covariates (covariates $Z_4$ to $Z_9$ in our application), then we choose to consider only unique timestamps, i.e. we aggregate orders of the same type occurring at the same timestamp.

\subsection{Estimation procedure of the two-step ratio model}

Following Sections \ref{sec:MarkedProcesses} and \ref{sec:QuasiLikelihoodEstimation}, estimation of the model defined at Equation \eqref{eq:TradesModel} can be carried out with multiple successive ratio models. In the first step, we consider the difference parameters $\theta^i_j = \vartheta^i_j - \vartheta^0_j, i\in\mathbb I\setminus\{0\}, j\in\mathbb J$ and the ratios $(i\in\mathbb I\setminus\{0\})$:
\begin{align}
	r^i(t,\theta) & = \frac{ \exp\left( \sum_{j\in\mathbb J} \vartheta^i_j X_ j(t)\right)}{ \sum_{i'\in\mathbb I} \exp\left( \sum_{j\in\mathbb J} \vartheta^{i'}_j X_ j(t)\right)}
	= \left[ \sum_{i'\in\mathbb I} \exp\left( \sum_{j\in\mathbb J} ( \theta^{i'}_j - \theta^i_j) X_ j(t)\right)\right]^{-1}.
\label{eq:ThetaRatioApplication}
\end{align}
The quasi-log-likelihood based on the observation on $[0,T]$ for this ratio model is defined at Equation \eqref{eq:ThetaQL}.
In the second step, we consider the ratios
\begin{align}
	p^{k_i}_i(t, \varrho^i) & = \frac{\exp\left(\sum_{j\in\mathbb J_i} \varrho^{i,k_i}_{j_i} Y^i_j(t) \right)}{\sum_{k'_i\in\mathbb K_i} \exp\left(\sum_{j\in\mathbb J_i} \varrho^{i,k'_i}_{j_i} Y^i_j(t) \right)}
	= \left[ \sum_{k'_i\in\mathbb K_i} \exp\left( \sum_{j\in\mathbb J_i} ( \varrho^{i,k'_i}_j - \varrho^{i,k_i}_{j_i}) X_ j(t)\right)\right]^{-1},
\label{eq:RhoRatioApplication}
\end{align}
and the associated quasi-log-likelihood of Equation \eqref{eq:RhoQL}.
Consistency and asymptotic normality of the quasi-maximum likelihood estimators are guaranteed by Theorem \ref{thm:ConvergenceResult}.

\subsection{In-sample model selection with QAIC}
\label{subsec:InSampleQAIC}

In this first application, we perform in-sample model selection to assess the relevance of the different possible sets of covariates.
For each stock and each trading day, we fix a set of covariates. We use the indices of the tested covariates to name the models: the model 146 is thus the model with covariates $(Z_1,Z_4,Z_6)$. If required, we estimate the parameters of all the Hawkes covariates and then compute the Hawkes covariates using these values. We finally fit three ratio models following the above procedure : one for the processes $(N^0, N^1)$ (signature of the marker orders), one for the processes $(N^{0,0}, N^{0,1})$ (aggressiveness of the bid market orders) and one for the processes $(N^{1,0}, N^{1,1})$ (aggressiveness of the ask market orders).

\begin{figure}
\centering
\includegraphics[width=0.75\textwidth]{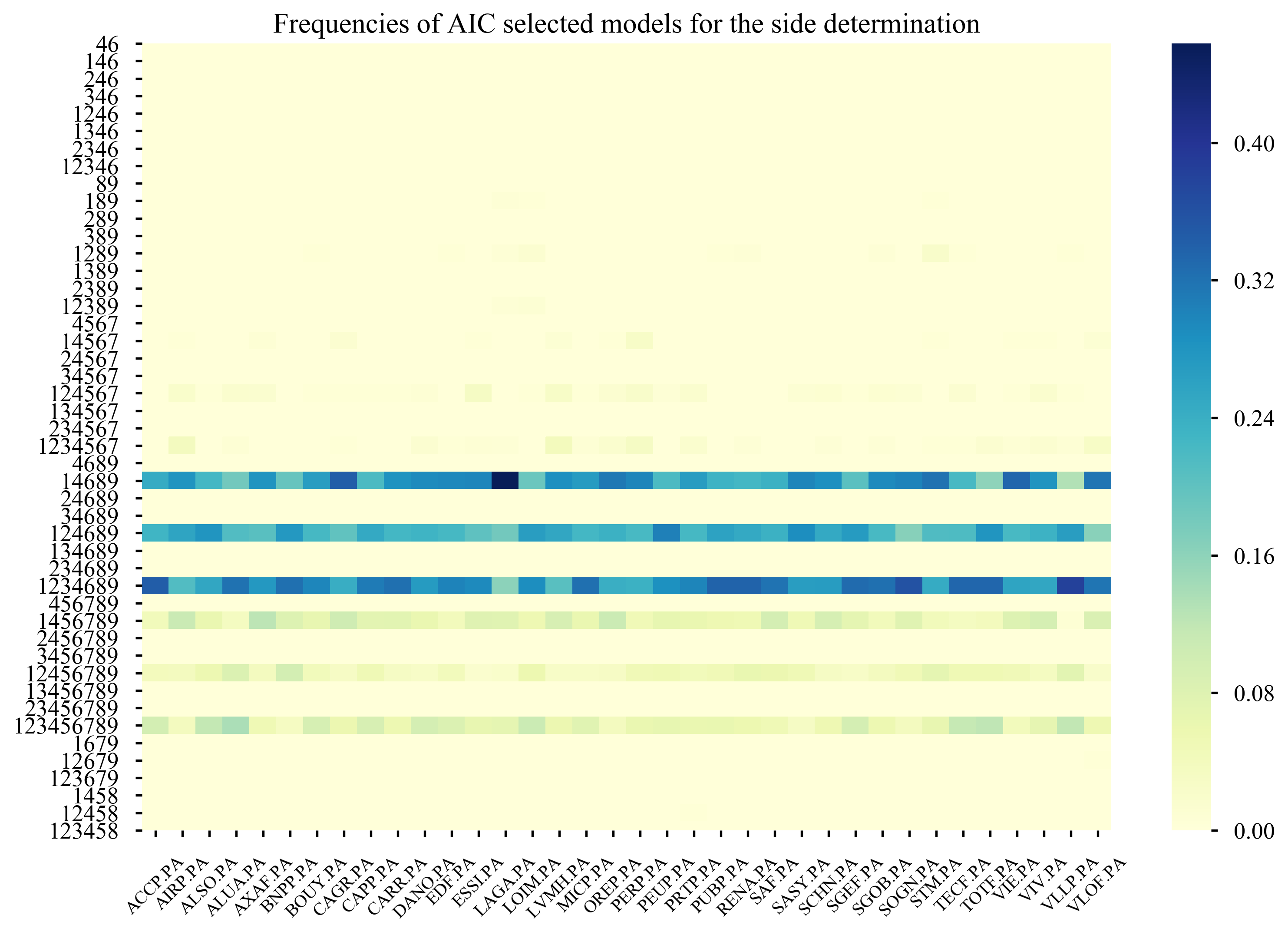}
\caption{Side of market orders - Frequency of selection of each model by the QAIC criterion, for each stock.}
\label{fig:AIC-SideSelection}
\end{figure}

\begin{figure}
\centering
\includegraphics[width=0.75\textwidth]{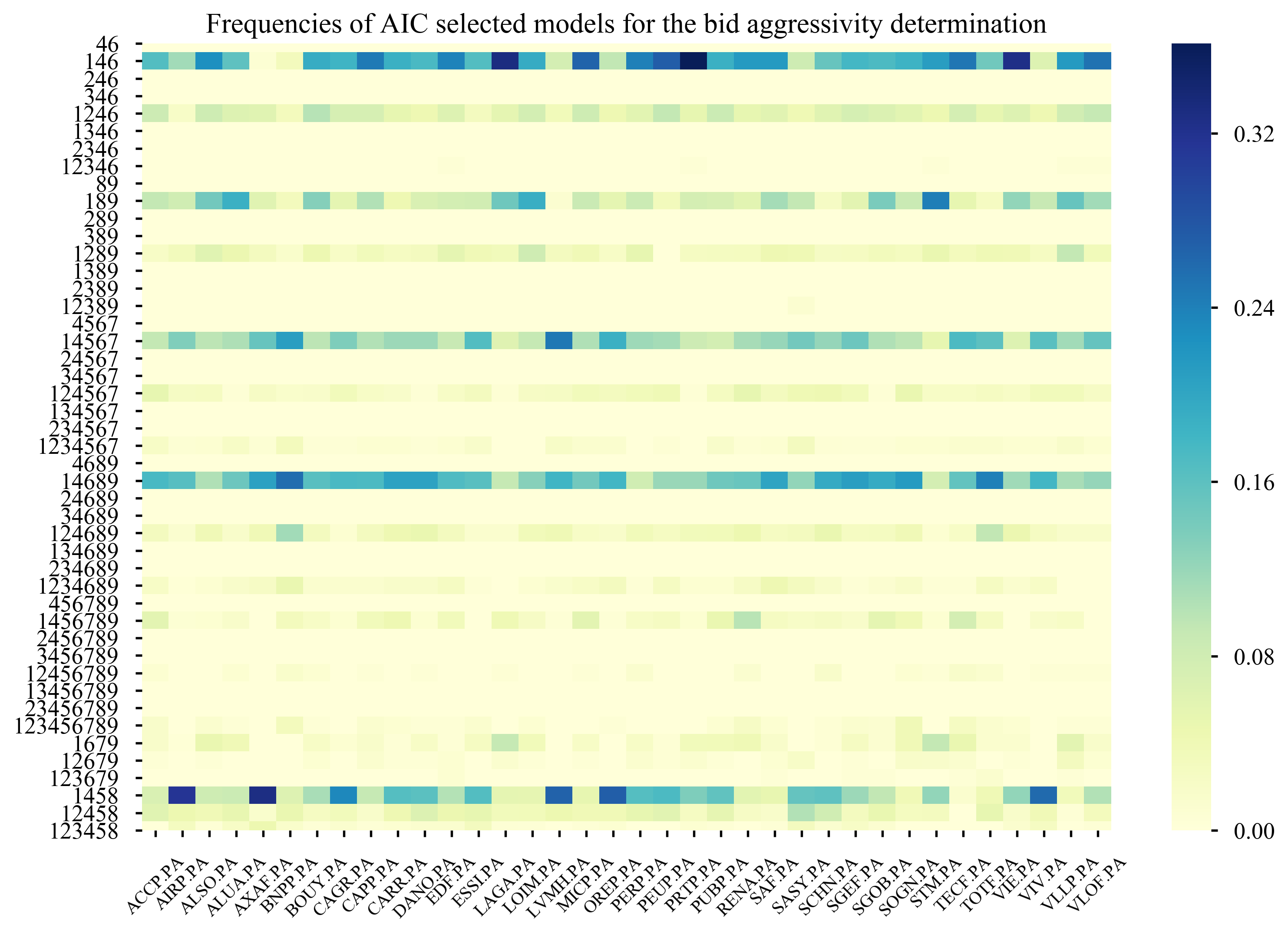}
\caption{Aggressiveness of bid market orders - Frequency of selection by the QAIC criterion of each model, for each stock.}
\label{fig:AIC-BidAggressivenessSelection}
\end{figure}

\begin{figure}
\centering
\includegraphics[width=0.75\textwidth]{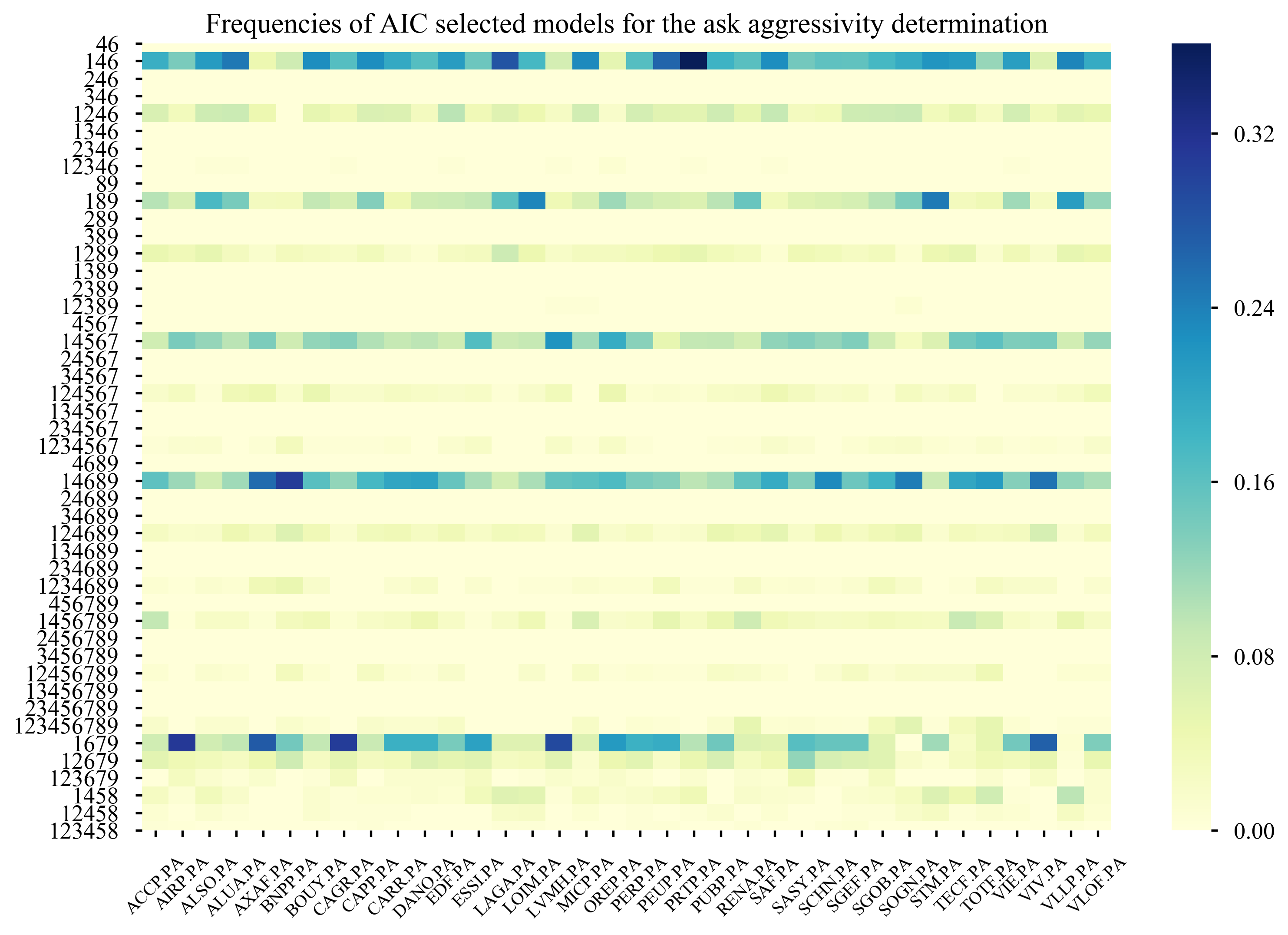}
\caption{Aggressiveness of ask market orders - Frequency of selection by the QAIC criterion of each model, for each stock.}
\label{fig:AIC-AskAggressivenessSelection}
\end{figure}

For each trading day, we then select the model minimizing the QAIC. For the ratio for the side determination, the criterion is
\begin{equation}
	-2 \mathbb H_T(\hat{\theta}^M_T)+2 |\mathbb J|,
\end{equation}
where $|\mathbb J|$ is the cardinality of the set of $\mathbb J$.
For the aggressiveness ratios, the criterion is
\begin{equation}
	-2 \mathbb H^{(i)}_T(\hat\varrho^{i})+2 |\mathbb J_i| \quad (i\in\mathbb I).
\end{equation}
We finally compute for each stock the frequencies of selection of different sets of covariates (i.e. the number of trading days in which a model is selected by QAIC over the total number of trading days in the sample for this stock).
Figures \ref{fig:AIC-SideSelection}, \ref{fig:AIC-BidAggressivenessSelection} and \ref{fig:AIC-AskAggressivenessSelection} plot the results as a model $\times$ stock heatmap for each of these three ratios.
For completeness, Tables \ref{table:AIC-side-ModelRanking}, \ref{table:AIC-aggB-ModelRanking} and \ref{table:AIC-aggA-ModelRanking} in Appendix \ref{sec:AICDetailedResults} list for each stock and each ratio model (side, bid aggressiveness, ask aggressiveness) the four most selected models (with frequency of selection).

For side determination, the models 14689, 124689 and 1234689 are the three most often chosen models: the selected model is among these three models approximately $80\%$ of the time in average across stocks. Imbalance, Hawkes covariates for bid and ask market orders, and Hawkes covariates for aggressive bid and ask market orders thus appear to be the most informative covariates. 

For aggressiveness determination, the model 146 is often selected. This is in line with intuition: imbalance is known to be a significant proxy for price change \citep{Lipton2013} and Hawkes covariates for aggressive bid and aggressive are specific to the targeted events. 
Note also that for several stocks, models with "symmetric" sets of covariates can also be chosen: for ask aggressiveness, 1679 is often selected, i.e. imbalance and all available ask Hawkes covariates ; symmetrically, 1458 is selected for ask aggressiveness, i.e. imbalance and all available bid Hawkes covariates.

\begin{figure}
\centering
\begin{tabular}{cc}
\includegraphics[width=0.45\textwidth]{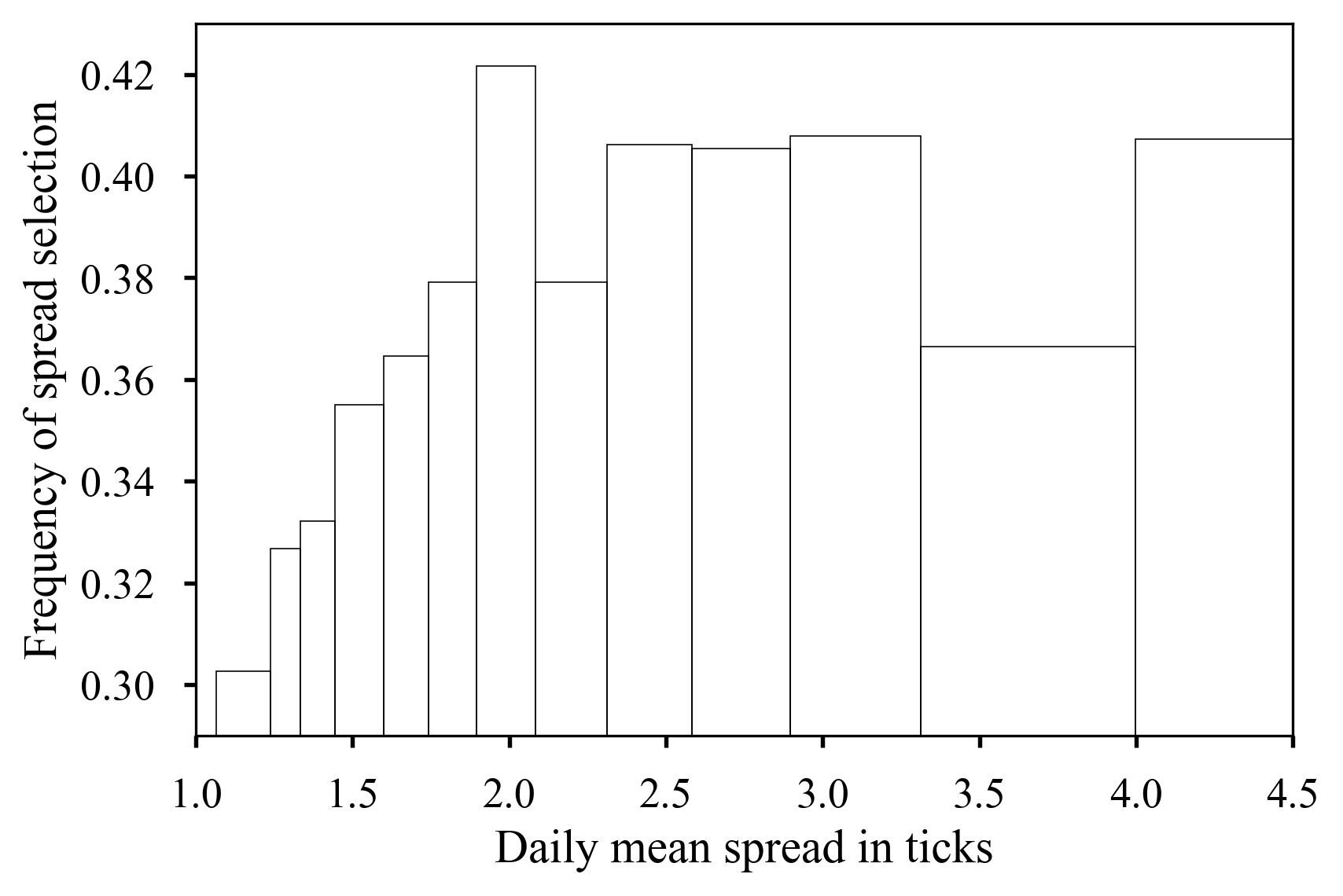}
&
\includegraphics[width=0.45\textwidth]{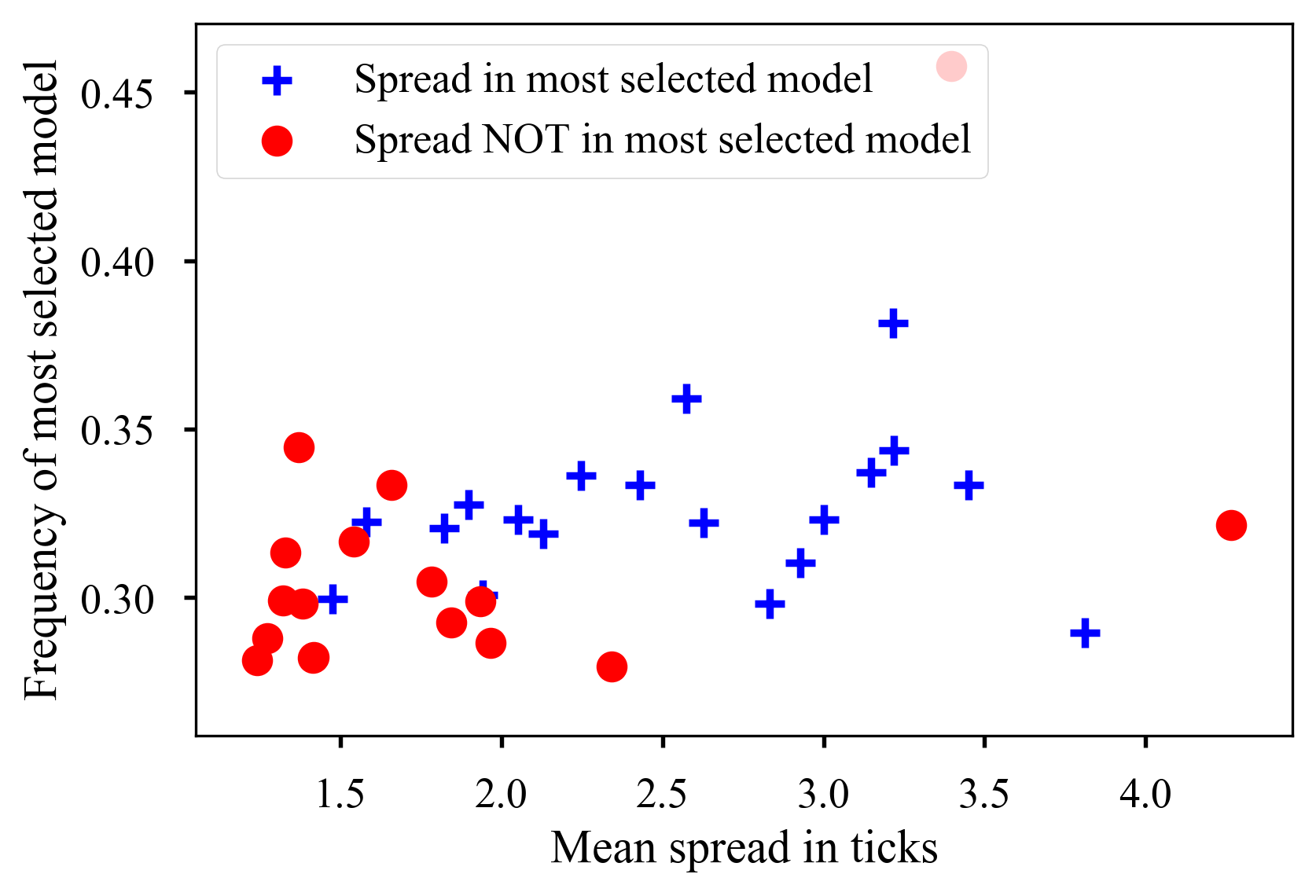}
\end{tabular}
\caption{Spread selection in the side ratio model as a function of the mean observed spread in ticks. Frequency of selection across all stocks and trading days (left) and frequency of selection for each stock (right).}
\label{fig:SpreadSelection}
\end{figure}

One may in particular observe that these results  confirm the primary role of the spread measured in ticks in the theory of financial microstructure. Stocks for which the observed spread is mostly equal to one tick are labeled 'large-tick stocks', implying that market participants are constrained by the price grid when submitting orders to the limit order book. Other stocks may be labeled 'small-tick stocks' \citep{Eisler2012}.
Using our sample, we compute the mean observed spread in ticks for each stock and each available trading day, and group these values in bins of equal sizes. Then inside each bin, we compute the frequency of selection of the covariate $X_3$ (signed spread) by QAIC for the ratio estimation of Equation \eqref{eq:ThetaRatioApplication}. Bar plot is provided in Figure \ref{fig:SpreadSelection} (left). We observe an increase of the frequency of the selection of the spread covariate when the mean observed spread increases from $1$ tick (its minimal possible value) to roughly $2$ ticks. For larger spread values, frequency then oscillates at high values. A break point might be searched between $1.5$ and $2$ ticks. This indicates that the significancy of covariates, especially the spread, is not the same for large-tick and small-tick stocks, and that even for small tick-stocks, dependency is not constant/uniform.

This observation is complemented on Figure \ref{fig:SpreadSelection} (right) by a cross-stock view of this phenomenon. For each stock, we plot the frequency of the most selected model, in blue if the spread is selected, in red if not. We observe that the spread covariate is nearly always in the most selected model for stocks with an observed mean spread larger than $2$ ticks. Recall that many microstructure models are developped for large-tick stocks, since assuming a constant spread equal to one tick often simplifies the analysis of the limit order book dynamics. Our observation advocates for the definition of specific microstructure models for small-tick stocks, taking into account the spread dynamics.

\subsection{Out-of-sample prediction performance}
\label{subsec:Prediction}

In this section, we use intensity and ratio models to predict the sign and aggressiveness of an incoming market order. For all tested models, the procedure is the following. On a given trading day, the model is fitted. Fitted parameters are then used \emph{on the following trading day} (available in the database) to compute the intensities (or ratios for ratio models), at all time. The type of an incoming event is then predicted to be the type of highest intensity or ratio. The exercise is theoretical in the sense that we assume that these computations are instantaneous, so that intensities or ratios are available at all times.

Recall the notation $N=(N^{i,k_i})_{i\in\{0,1\}, k_i\in\{0,1\}}$ for the four-dimensional point process counting bid aggressive market orders, bid non-aggressive market orders, ask aggressive market orders and ask non-aggressive market orders.
We use two benchmark models.

The first benchmark model is the Hawkes model. Here, $N$ is assumed to be a four-dimensional Hawkes process with a single exponential kernel. In vector notation, the intensity is written
\begin{equation*}
	\begin{pmatrix}
		\lambda_H^{0,0}(t) \\
		\lambda_H^{0,1}(t) \\
		\lambda_H^{1,0}(t) \\
		\lambda_H^{1,1}(t) 
	\end{pmatrix}
	= 
	\begin{pmatrix}
		\mu^{0,0} \\
		\mu^{0,1} \\
		\mu^{1,0} \\
		\mu^{1,1} 
	\end{pmatrix}
	+ \int_{0}^t 
	\begin{pmatrix}
		\ddots \;\;\;\;\;\; \vdots  \;\;\;\;\;\; \iddots \\
		\alpha_{(i,k_i),(j,k_j)} e^{-\beta_{(i,k_i),(j,k_j)}(t-s)}
		\\ 	\iddots  \;\;\;\;\;\; \vdots  \;\;\;\;\;\; \ddots \\
	\end{pmatrix}		
	\cdot
	\begin{pmatrix}
		dN^{0,0}(s) \\
		dN^{0,1}(s) \\
		dN^{1,0}(s) \\
		dN^{1,1}(s) 
	\end{pmatrix}
\end{equation*}
Estimation and ratio computation can be found in e.g., \cite{Bowsher2007,MuniTokePomponio2012}. This model is labeled 'Hawkes'.

The second benchmark model is the four-dimensional ratio model without marks \citep{MuniTokeYoshida2019}. In this model, the intensity of the counting process $(i,k_i)$ is 
\begin{equation*}
	\lambda_{R}^{i,k_i}(t) = \lambda_{0,R}(t) \exp\bigg(\sum_{j\in\bbJ}
\vartheta^{i,k_i}_j X_j(t) \bigg),
\end{equation*}
with some unobserved baseline intensity $\lambda_{0,R}(t)$. Given the previous observations, we choose the set of covariates $(Z_1,Z_4,Z_6,Z_8,Z_9)$ for this benchmark. It is natural to choose these covariates (imbalance, Hawkes for aggressive orders and Hawkes for all orders) given the results on model selection of Section \ref{subsec:InSampleQAIC}. Estimation and ratio computation are detailed in \cite{MuniTokeYoshida2019}. This model is labeled 'Ratio-14689'.

These two benchmarks are used to assess the performances of two marked ratio models (or two-step ratio models) described in this paper. The first marked ratio model uses the covariates $(Z_4,Z_5,Z_6,Z_7)$ for both steps. These covariates are based on the Hawkes processes of the benchmark Hawkes model. The second marked ratio model uses the covariates $(Z_1,Z_4,Z_6,Z_8,Z_9)$ for the first-step ratio (side determination) and $(Z_1,Z_4,Z_6)$ for both second-step ratios (bid and ask aggressiveness). Again, these choices are natural given the results on model selection of Section \ref{subsec:InSampleQAIC}. These models are labeled 'MarkedRatio-4567-4567-4567' and 'MarkedRatio-14689-146-146' respectively.

\begin{figure}
\centering
\includegraphics[width=0.8\textwidth]{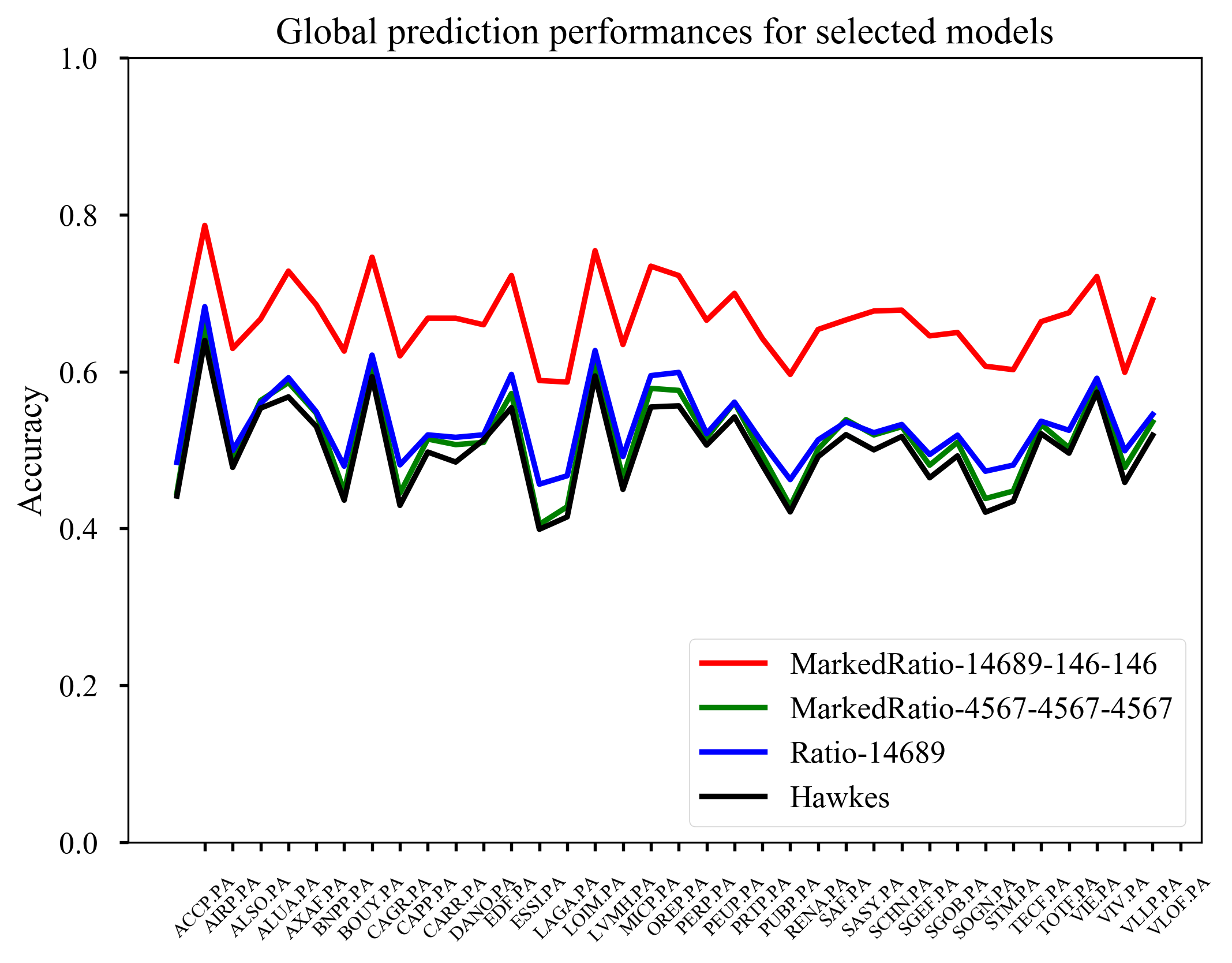}
\caption{Out-of-sample prediction performances for the benchmark models and the marked ratio models. Label explanation is in the text.}
\label{fig:PredictionPerformances}
\end{figure}

\begin{figure}
\centering
\begin{tabular}{cc}
\includegraphics[width=0.45\textwidth]{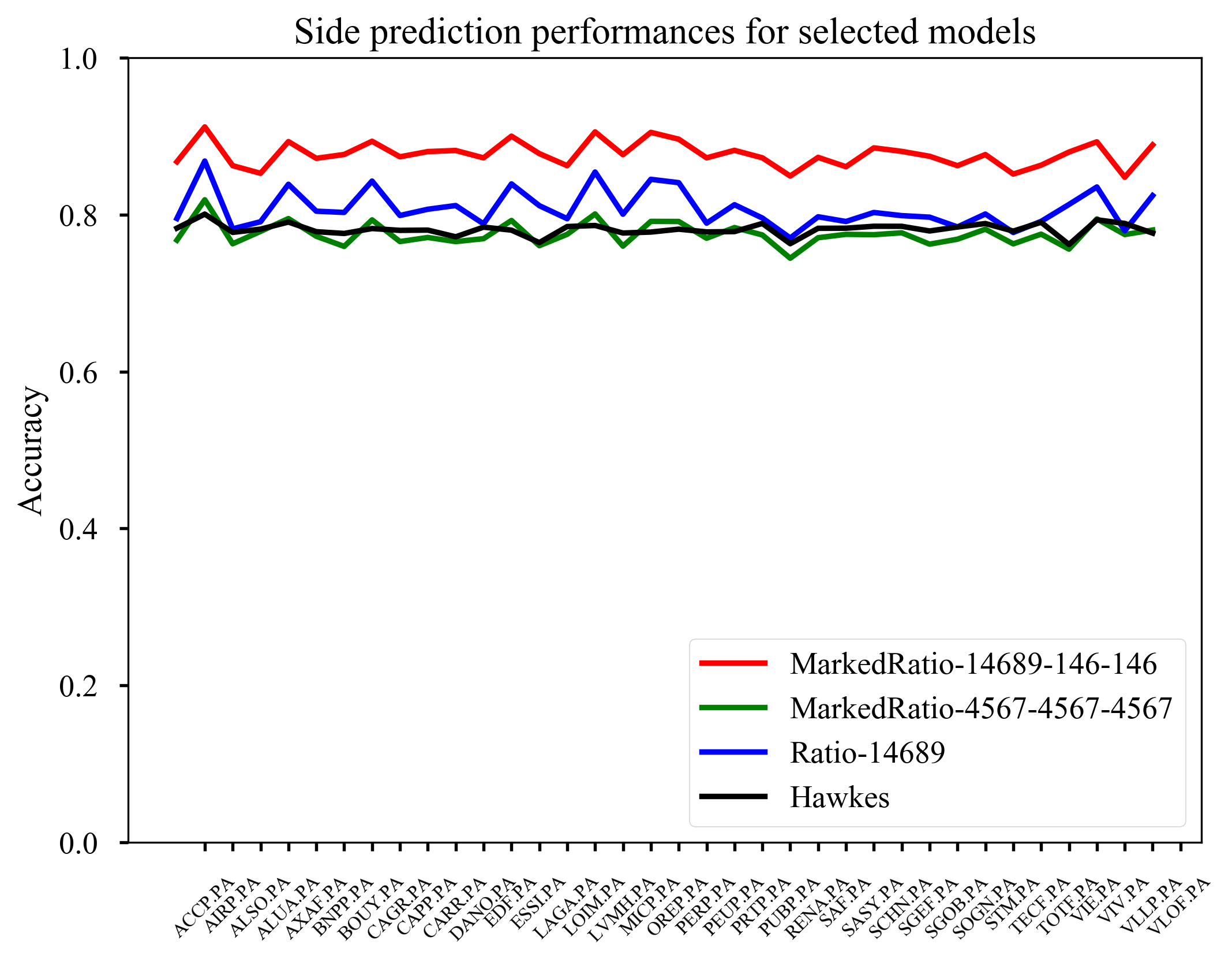}
&
\includegraphics[width=0.45\textwidth]{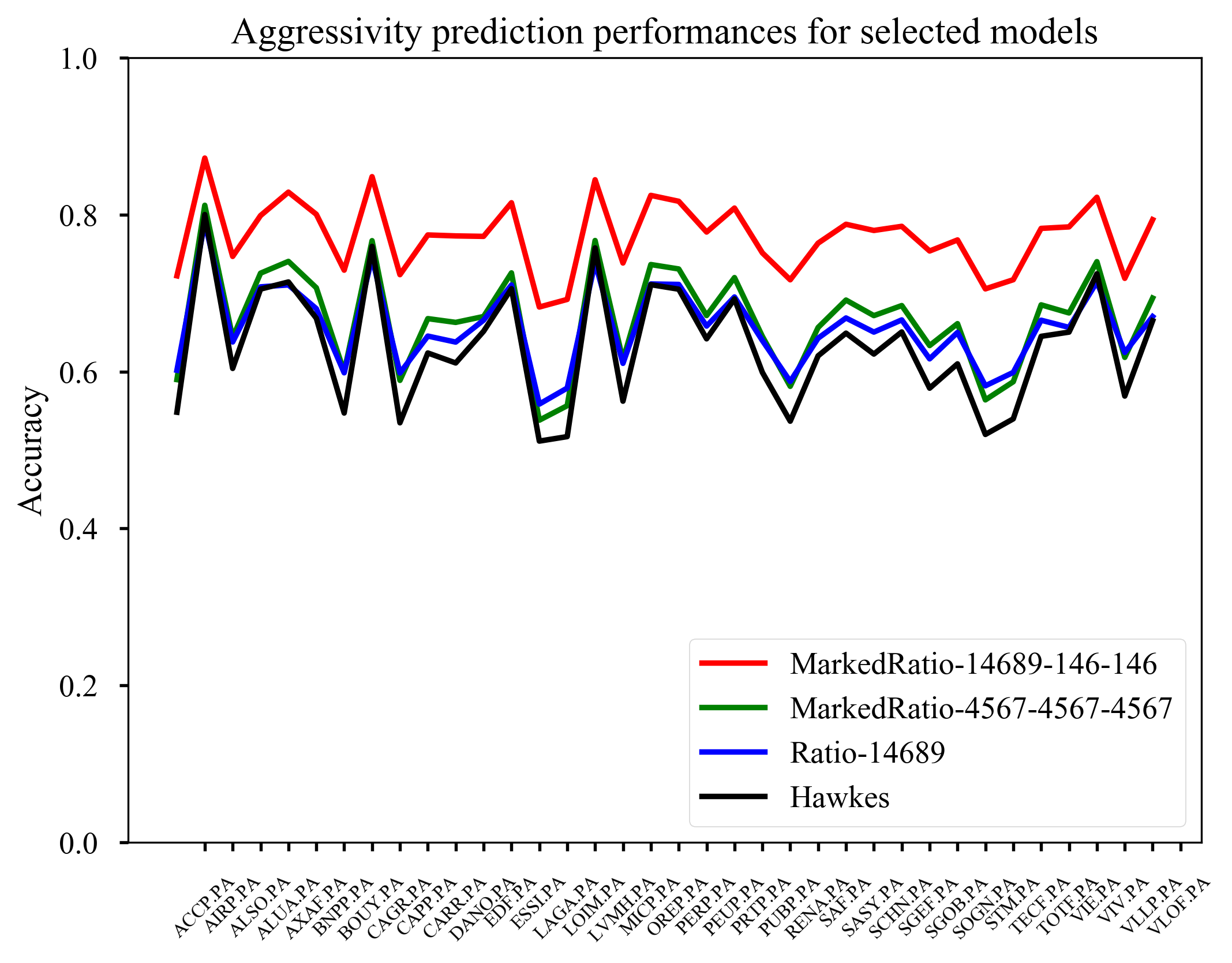}
\end{tabular}
\caption{Out-of-sample partial prediction performances for the side prediction (left) and aggressiveness prediction (right), for the benchmark models and the marked ratio models. Label explanation is in the text.}
\label{fig:PartialPredictionPerformances}
\end{figure}

Figure \ref{fig:PredictionPerformances} plots the results for each stock for the two benchmark models and the two marked ratio models.
For completeness, the partial performances for side determination and aggressiveness determination of the trades are provided on Figure \ref{fig:PartialPredictionPerformances}.
Finally, Table \ref{table:PredictionPerformances} lists the partial and global prediction performances of these models averaged across stocks.
\begin{table}
\centering
\begin{tabular}{|c|cccc|}
\hline
\multirow{2}{*}{Accuracy} & \multirow{2}{*}{Hawkes} & \multirow{2}{*}{Ratio-14689} &  MarkedRatio &  MarkedRatio
\\
&  &  &  4567-4567-4567 &  14689-146-146
\\ \hline
partial - side & 0.781 & 0.808 & 0.776 & \textbf{0.877}
\\
partial - agg. & 0.634 & 0.658 & 0.668 & \textbf{0.774}
\\ \hline
global & 0.503 & 0.533 & 0.516 & \textbf{0.667}
\\ \hline
\end{tabular}
\caption{Prediction performances of selected models averaged across stocks. Side accuracy gives the fraction of correctly signed trades. Aggressiveness accuracy gives the proportion of trade with a correctly predicted accuracy. Global accuracy gives the fraction of orders with correctly predicted side and aggressiveness.}
\label{table:PredictionPerformances}
\end{table}
The benchmark Hawkes model correctly predicts the sign and aggressiveness of an incoming with an accuracy in the range $[40\%,60\%]$ for all stocks, with a $50\%$ average. The marked ratio model with only Hawkes parameters ('MarkedRatio-4567-4567-4567') and no dependency on the state of the limit order book actually reproduces closely these performances. The non-marked ratio model 'Ratio-14689' improves slightly the global performances of the two previous models. When looking at the partial accuracies, we observe that this improvement is mainly due to a better side prediction. Finally, the 'MarkedRatio-14689-146-146', which appeared to be in average the best model with respect to the QAIC selection, results strongly outperforms all other models. The global accuracy is in the range $[60\%,80\%]$ for all stocks, with a $67\%$ average, i.e. we are theoretically able to correctly predict both the sign and aggressiveness of an incoming market order two times out of three.

These results show that the two-step ratio model for marked point processes is a significant improvement to existing intensity models. As in the standard ratio model of \cite{MuniTokeYoshida2019}, this provides an easy way to have both clustering and state-dependency. However, it is important to note that the two-step ratio strongly improves the performance of the standard ratio model in multidimensional setting. In this example, flexibility in the choice of covariates allows for precise model selection for both sign and aggressiveness.

\section*{Acknowledgements}
This work was in part supported by Japan Science and Technology Agency CREST JPMJCR14D7; 
Japan Society for the Promotion of Science Grants-in-Aid for Scientific Research No. 17H01702 (Scientific Research) and by a Cooperative Research Program of the Institute of Statistical Mathematics.

\bibliographystyle{authordate1}
\bibliography{marked}

\appendix

\section{Proof of Theorem \ref{201911220134}}\label{201911220137}
\label{sec:TheoremProof}
The convergence given in Theorem \ref{201911220134} can be obtained by 
the quasi-likelihood analysis, which we recall in Section \ref{201911221017}. 
We will apply Theorems \ref{201911221015} and \ref{201912011335} 
in Section \ref{201911221017} to the double ratio model. 
In the present situation, the scaling factor is $b_T=T$, 
the joint parameter $(\theta,\rho)$ is for $\theta$ in Section \ref{201911221017}, 
and the dimension of the full parameter space is $\check{\sfp}$ 
in place of $\sfp$ of Section \ref{201911221017}. 
Fix a set of values of parameters 
$(\alpha,\beta_1,\beta_2,\rho, \rho_1,\rho_2)$
so that Condition $[L1]$ (Section \ref{201911221017}) is met with $\rho=2$. 

\subsection{Score functions and a central limit theorem}
The score function for $\rho^i$ is given by 
\beas 
F^{(i)}_T(\rho^i) &=& \partial_{\rho^i}\bbH_T^{(i)}(\rho^i)
\yeq
\sum_{k_i\in\bbK_i}\int_0^T\partial_{\rho^i}\log q^{k_i}_i(t,\rho^i)dN^{i,k_i}_t.
\eeas
Then 
\bea\label{201911220602}
F^{(i)}_T(\rho^i)
&=&
\sum_{k_i\in\bbK_i}\int_0^T 
\big(
1_{\{k_i\}}(\cdot)-q^\flat_i(t,\rho^i)\big)\otimes\bbY^i(t)dN^{i,k_i}_t
\eea
where $q^\flat_i(t,\rho^i)=(q_i^{k_i}(t,\rho^i))_{k_i\in\bbK_{i,0}}$ and $\bbY^i(t)=(Y^i_{j_i}(t))_{j_i\in\bbJ_i}$. 
By some calculus, we see 
\bea\label{201911301926} 
F^{(i)}_T:=F^{(i)}_T((\rho^i)^*)
&=&
\sum_{k_i\in\bbK_i}\int_0^T 
\big(
1_{\{k_i\}}(\cdot)-q^\flat_i(t,(\rho^i)^*)\big)\otimes\bbY^i(t)d\tilde{N}^{i,k_i}_t
\eea
We are assuming that the counting processes $N^{i,k_i}$ ($i\in\bbI;\>j_i\in\bbK_i$) have no common jumps. 
Then the $\sfp_i\times\sfp_{i'}$ matrix valued process
\bea\label{201912010110} 
\langle F^{(i)},F^{(i')}\rangle_T &=& 0\quad(i,i'\in\bbI,\>i\not=i')
\eea
and 
\beas 
\langle F^{(i)}\rangle_T
&=&
\sum_{k\in\bbK_i}\int_0^T\bigg\{1_{\{k_i\}}(\cdot)-q^\flat_i(t,(\rho^i)^*)\big)\otimes\bbY^i(t)\bigg\}^{\otimes2}
r^i(t,\theta^*)\Lambda(\lambda_0(t),\bbX(t))q^{k_i}_i(t,(\rho^i)^*)dt
\\&=&
\int_0^T{\sf V}^i_0(\bbY^i(t),(\rho^i)^*)\otimes(\bbY^i(t))^{\otimes2}
\>\Lambda(\lambda_0(t),\bbX(t))r^i(t,\theta^*)dt
\quad(i\in\bbI)
\eeas
Therefore, the mixing property $[M2]$ gives the convergence
\bea\label{201912010111}
T^{-1}\langle F^{(i)}\rangle_T
&\to^p&
\Gamma^{(i)}((\rho^i)^*)=
E\bigg[{\sf V}^i_0(\bbY^i(0),(\rho_i)^*)\otimes(\bbY^i(0))^{\otimes2}
\>\Lambda(\lambda_0(t),\bbX(0))r^i(0,\theta^*)\bigg]
\eea
as $T\to\infty$, with the aid of $[M1]$. 

The score function for $\theta$ is the $\sfp$-dimensional process 
\bea\label{201912010409} 
F_T(\theta) &=& \partial_{\theta}\bbH_T(\theta)
\yeq 
\sum_{i\in\bbI}\int_0^T\partial_{\theta}\log r^i(t,\theta)dN^i_t
\nn\\&=&
\sum_{i\in\bbI}\int_0^T
\big(1_{\{i\}}(\cdot)-r^\flat(t,\theta)\big)\otimes\bbX(t)dN^i_t.
\eea
where $r^\flat(t,\theta)=(r^i(t,\theta))_{i\in\bbI_0}$. 
Evaluated at $\theta^*$, 
\bea\label{201911301925}
F_T
&=&
F_T(\theta^*)
\yeq
\sum_{i\in\bbI}\int_0^T\big(1_{\{i\}}(\cdot)-r^\flat(t,\theta^*)\big)\otimes \bbX(t)d\tilde{N}^i_t
\nn\\&=&
\sum_{i\in\bbI}\sum_{k_i\in\bbK_i}\int_0^T\big(1_{\{i\}}(\cdot)-r^\flat(t,\theta^*)\big)\otimes \bbX(t)d\tilde{N}^{i,k_i}_t.
\eea
Then, the $\sfp\times\sfp$ matrix valued process $\langle F\rangle$ has the expression
\beas 
\langle F\rangle_T
&=&
\sum_{i\in\bbI}\sum_{k_i\in\bbK_i}\int_0^T\big(1_{\{i\}}(\cdot)-r^\flat(t,\theta^*)\big)^{\otimes2}
\otimes \bbX(t)^{\otimes2}
r^i(t,\theta^*)\Lambda(\lambda_0(t),\bbX(t))q^{k_i}_i(t,(\rho_i)^*)dt
\\&=&
\sum_{i\in\bbI}\int_0^T\big(1_{\{i\}}(\cdot)-r^\flat(t,\theta^*)\big)^{\otimes2}
\otimes \bbX(t)^{\otimes2}
r^i(t,\theta^*)\Lambda(\lambda_0(t),\bbX(t))dt
\\&=&
\int_0^T{\sf V}_0(\bbX(t))\otimes \bbX(t)^{\otimes2}\Lambda(\lambda_0(t),\bbX(t))dt.
\eeas
Then the mixing property $[M2]$ provides the convergence 
\bea\label{201912010128} 
T^{-1}\langle F\rangle_T
&\to^p&
\Gamma(\theta^*)
\yeq
E\bigg[\bigg({\sf V}_0(\mathbb X(0))\otimes \mathbb X(0)^{\otimes2}\bigg)
\Lambda(\lambda_0(0),\mathbb X(0))\bigg]
\eea
as $T\to\infty$. 

For $i\in\bbI$, 
\bea\label{201912010129}
\langle F,F^{(i)}\rangle_T
&=&
\sum_{k_i\in\bbK_i}\int_0^T
\big(1_{\{i\}}(\cdot)-r^\flat(t,\theta^*)\big)\otimes 
\big(1_{\{k_i\}}(\cdot)-q^\flat_i(t,(\rho^i)^*)\big)
\otimes\bbX(t)\otimes\bbY^i(t)
\nn\\&&\qquad\qquad\times 
r^i(t,\theta^*)\Lambda(\lambda_0(t),\bbX(t))q^{k_i}_i(t,(\rho_i)^*)dt
\nn\\&=&
0
\eea
since 
\beas 
\sum_{k_i\in\bbK_i}\big(1_{\{k_i\}}(\cdot)-q^\flat_i(t,(\rho^i)^*)\big)q^{k_i}_i(t,(\rho_i)^*)
&=&
0.
\eeas

The full information matrix is 
the $\check{\sfp}\times\check{\sfp}$ block diagonal matrix
\beas 
\Gamma\yeq\Gamma(\theta^*,\rho^*)
&=&
\text{diag}\big[\Gamma(\theta^*),\Gamma^0((\rho^0)^*),
\Gamma^1((\rho^1)^*),...,\Gamma^{\bar{i}}((\rho^{\bar{i}})^*)\big]
\eeas
Let $\Delta_T=T^{-1/2}\big(F_T,(F^{(i)}_T)_{i\in\bbI}\big)$. 
Now, by the martingale central limit theorem, it is easy to obtain the convergence 
\beas 
\Delta_T &\to^d& \Gamma^{1/2}\zeta\quad(T\to\infty)
\eeas
where $\zeta$ is a $\check{\sfp}$-dimensional standard Gaussian random vector. 
The joint convergence $(\Delta_T,\Gamma)\to^d(\Gamma^{1/2}\zeta,\Gamma)$ is obvious 
since $\Gamma$ is deterministic.

\subsection{Condition $[L4]$}
According to (\ref{201912010222}), we 
define the random field $\bbY_T:\Omega\times\overline{\Theta}\times\overline{\calr}\to\bbR$ by 
\beas
\bbY_T(\theta,\rho) = T^{-1}\big(\bbH_T(\theta,\rho)-\bbH_T(\theta^*,\rho^*)\big)
\eeas 
for $\bbH_T(\theta,\rho)$ given in (\ref{201909120239}). 
From the expression (\ref{201911301554}) of $\bbH_T(\theta,\rho)$, we have 
\beas 
T^{-1}\bbH_T(\theta,\rho)
&=&
T^{-1}\sum_{i\in\bbI}\sum_{k_i\in\bbK_i}
\int_0^T\log \big(r^i(t,\theta)q^{k_i}_i(t,\rho^i)\big) dN^{i,k_i}_t
\\&=&
T^{-1}\sum_{i\in\bbI}\sum_{k_i\in\bbK_i}
\int_0^T\log \big(r^i(t,\theta)q^{k_i}_i(t,\rho^i)\big)d\tilde{N}^{i,k_i}_t
\\&&
+T^{-1}\sum_{i\in\bbI}\sum_{k_i\in\bbK_i}\int_0^T
\big\{\log \big(r^i(t,\theta)q^{k_i}_i(t,\rho^i) \big\}
\lambda_0(t)\exp\bigg(\sum_{j\in\bbJ}
(\vartheta^*)^i_jX_j(t)\bigg)\>p_i^{k_i}(t,(\varrho^*)^{i,k_i})dt.
\eeas
By definition, 
\beas 
\big|\partial_{(\theta,\rho)}^\ell\log \big(r^i(t,\theta)q^{k_i}_i(t,\rho^i)\big)\big|
&\leq&
C\bigg(1+\sum_{j\in\bbJ}|X_j(t)|+\sum_{i\in\bbI}\sum_{j_i\in\bbJ_i}|Y^{k_i}_{j_i}(t)|\bigg)
\quad(\ell=0,1)
\eeas
where $C$ is a constant depending on the diameters of $\Theta$ and $\calr$. 
Therefore, under Condition $[M1]$,  
\beas &&
E\bigg[\bigg|\partial_{(\theta,\rho)}^\ell T^{-1/2}
\int_0^T\log \big(r^i(t,\theta)q^{k_i}_i(t,\rho^i)\big)d\tilde{N}^{i,k_i}_t\bigg|^{2^k}\bigg]
\\&\simleq&
E\bigg[\bigg(T^{-1}
\int_0^T\big|\partial_{(\theta,\rho)}^\ell \log \big(r^i(t,\theta)q^{k_i}_i(t,\rho^i)\big)\big|^2
dN^{i,k_i}_t\bigg)^{2^{(k-1)}}\bigg]
\\&\simleq&
E\bigg[T^{-1}
\int_0^T\big|\partial_{(\theta,\rho)}^\ell \log \big(r^i(t,\theta)q^{k_i}_i(t,\rho^i)\big)\big|^{2^k}
\lambda^{i,k_i}(t,(\vartheta^i)^*,(\varrho^{i,k_i})^*)dt\bigg]
\\&&
+T^{-2^{k-2}}E\bigg[\bigg(T^{-1/2}
\int_0^T\big|\partial_{(\theta,\rho)}^\ell \log \big(r^i(t,\theta)q^{k_i}_i(t,\rho^i)\big)\big|^2
d\tilde{N}^{i,k_i}_t\bigg)^{2^{(k-1)}}\bigg]
\\&=&
O(1)+T^{-2^{k-2}}E\bigg[\bigg(T^{-1/2}
\int_0^T\big|\partial_{(\theta,\rho)}^\ell \log \big(r^i(t,\theta)q^{k_i}_i(t,\rho^i)\big)\big|^2
d\tilde{N}^{i,k_i}_t\bigg)^{2^{(k-1)}}\bigg]
\eeas
for $k\in\bbN$, 
where the constant appearing at each $\simleq$ depends only on $\check{\sfp}$, $k$ and 
the constant of the Burkholder-Davis-Gundy inequality. 
By induction, we obtain 
\bea\label{201912010230} 
\sup_{(\theta,\rho)\in\Theta\times\calr}\sup_{T\geq1}\bigg\|\partial_{(\theta,\rho)}^\ell T^{-1/2}
\int_0^T\log \big(r^i(t,\theta)q^{k_i}_i(t,\rho^i)\big)d\tilde{N}^{i,k_i}_t\bigg\|_p
&<&
\infty
\eea
for every $p>1$ and $\ell\in\{0,1\}$. 
Then Sobolev's inequality gives 
\bea\label{201911301717}
\sup_{T\geq1}
\bigg\|
\sup_{(\theta,\rho)\in\Theta\times\calr}
\bigg|T^{-1/2}\int_0^T\log \big(r^i(t,\theta)q^{k_i}_i(t,\rho^i)\big)d\tilde{N}^{i,k_i}_t\bigg|\>\bigg\|_p
&<&
\infty
\eea
for every $p>1$. 

Let 
\beas 
\Phi(t,\theta,\rho)
&=&
\sum_{i\in\bbI}
\sum_{k_i\in\bbK_i}\bigg\{r^i(t,\theta^*)p_i^{k_i}(t,(\varrho^*)^{i,k_i})
\log\frac{r^i(t,\theta)q^i(t,k_i,\rho^i)}{r^i(t,\theta^*)q_i^{k_i}(t,(\rho^*)^{i,k_i})}\bigg\}
\\&&\qquad\times
\lambda_0(t)\sum_{i'\in\bbI}\exp\bigg(\sum_{j\in\bbJ}
(\vartheta^*)^{i'}_jX_j(t)\bigg).
\eeas
Then Conditions $[M1]$ and $[M2]$ imply 
\beas 
\sup_{(\theta,\rho)\in\Theta\times\calr}
\sup_{T\geq1}\bigg\|T^{-1/2}\int_0^T\partial_{(\theta,\rho)}^\ell\big(\Phi(t,\theta,\rho)-E[\Phi(t,\theta,\rho)]\big)dt\bigg\|_p
&<&
\infty
\eeas
for every $p>1$ and $\ell\in\{0,1\}$. 
This entails 
\bea\label{201911301753} 
\sup_{T\geq1}\bigg\|T^{1/2}\sup_{(\theta,\rho)\in\Theta\times\calr}
\bigg|T^{-1}\int_0^T\Phi(t,\theta,\rho)dt-E[\Phi(t,\theta,\rho)]\bigg|\bigg\|_p
&<&
\infty
\eea
for every $p>1$. 

Combining (\ref{201911301753}) with (\ref{201911301717}), we obtain 
\bea\label{201912010433}
\sup_{T\geq1}E\bigg[\bigg(T^{1/2}\sup_{(\theta,\rho)\in\Theta\times\calr}
\big|\bbY_T(\theta,\rho)-\bbY(\theta,\rho)\big|\bigg)^p\bigg]
&<&
\infty
\eea
for every $p>1$, if we set 
\beas 
\bbY(\theta,\rho)
&=&
E\bigg[\sum_{i\in\bbI}
\sum_{k_i\in\bbK_i}\bigg\{r^i(0,\theta^*)p_i^{k_i}(0,(\varrho^*)^{i,k_i})
\log\frac{r^i(0,\theta)q^i(0,k_i,\rho^i)}{r^i(0,\theta^*)q_i^{k_i}(0,(\rho^*)^{i,k_i})}\bigg\}
\\&&\qquad\times
\lambda_0(0)\sum_{i'\in\bbI}\exp\bigg(\sum_{j\in\bbJ}
(\vartheta^*)^{i'}_jX_j(0)\bigg)\bigg]. 
\eeas
This verifies Condition $[L4]$(ii).

As (\ref{201912010242}), we define $\Gamma_T(\theta,\rho)$ by \label{Gamma}
\beas 
\Gamma_T(\theta,\rho)
&=&
-T^{-1}\partial_{(\theta,\rho)}^2\bbH_T(\theta,\rho).
\eeas
From (\ref{201911220602}), 
\beas
\partial_{\rho^i}^2\bbH_T^{(i)}(\rho^i)
&=&
-\sum_{k_i\in\bbK_i}\int_0^T 
\partial_{\rho^i}q^\flat_i(t,\rho^i)\otimes\bbY^i(t)dN^{i,k_i}_t.
\eeas
More precisely, 
\beas
\partial_{\rho^{i,k_i}_{j_i}}\partial_{\rho^{i,k_i'}_{j_i'}}\bbH_T^{(i)}(\rho^i)
&=&
-\sum_{k_i''\in\bbK_i}\int_0^T \bigg\{
1_{\{k_i=k_i'\}}q^{k_i}_i(t,\rho^i)
-q^{k_i}_i(t,\rho^i)q^{k_i'}_i(t,\rho^i)
\bigg\}\bbY^i_{j_i}(t)\bbY^i_{j_i'}(t)dN^{i,k_i''}_t
\\&=&
-\sum_{k_i''\in\bbK_i}\int_0^T \bigg\{
1_{\{k_i=k_i'\}}q^{k_i}_i(t,\rho^i)
-q^{k_i}_i(t,\rho^i)q^{k_i'}_i(t,\rho^i)
\bigg\}\bbY^i_{j_i}(t)\bbY^i_{j_i'}(t)d\tilde{N}^{i,k_i''}_t
\\&&
-\int_0^T \bigg\{
1_{\{k_i=k_i'\}}q^{k_i}_i(t,\rho^i)
-q^{k_i}_i(t,\rho^i)q^{k_i'}_i(t,\rho^i)
\bigg\}\bbY^i_{j_i}(t)\bbY^i_{j_i'}(t)
\\&&\qquad\qquad\qquad\times
r^i(t,\theta^*)\Lambda(\lambda_0(t),\bbX(t))dt
\\&=&
-\sum_{k_i''\in\bbK_i}\int_0^T {\sf V}_0^i(\bbY^i(t),\rho^i)_{k_i,k_i'}\bbY^i_{j_i}(t)\bbY^i_{j_i'}(t)d\tilde{N}^{i,k_i''}_t
\\&&
-\int_0^T 
{\sf V}_0^i(\bbY^i(t),\rho^i)_{k_i,k_i'}
\bbY^i_{j_i}(t)\bbY^i_{j_i'}(t)
\Lambda(\lambda_0(t),\bbX(t))r^i(t,\theta^*)dt
\eeas
for $k_i,k_i'\in\bbK_{i,0}$, $j_i,j_i'\in\bbJ_i$ and $i\in\bbI$, 
where (\ref{201912010321}) was used. 
Similarly, from (\ref{201912010409}), 
\beas 
\partial_\theta^2\bbH_T(\theta)
&=&
-\sum_{i\in\bbI}\int_0^T\partial_\theta r^\flat(t,\theta)\otimes\bbX(t)dN^i_t,
\eeas
equivalently, 
\beas 
\partial_{\theta^i_j}\partial_{\theta^{i'}_{j'}}\bbH_T(\theta)
&=&
-\sum_{i''\in\bbI}\int_0^T{\sf V}_0(\bbX(t),\theta)_{i,i'}X_j(t)X_{j'}(t)d\tilde{N}^{i''}_t
\\&&%
-\int_0^T{\sf V}_0(\bbX(t),\theta)_{i,i'}X_j(t)X_{j'}(t)\Lambda(\lambda_0(t),\bbX(t))dt
\eeas
for $i,i'\in\bbI_0$ and $j,j'\in\bbJ$. 
Obviously, 
\beas 
\partial_{\theta}\partial_{\rho^i}\bbH_T^{(i)}(\rho^i)\yeq0
\quad\text{and}\quad
\partial_{\rho^{i'}}\partial_{\rho^i}\bbH_T^{(i)}(\rho^i)
&=&
0
\quad(i',i\in\bbI:\>i'\not=i)
\eeas
In a way similar to the derivation of (\ref{201912010433}), as a matter of fact it is easier, 
we can show 
\beas 
\sup_{T\geq1}E\big[\big(T^{1/2}|\Gamma_T(\theta^*,\rho^*)-\Gamma|\big)^p\big]
&<&
\infty
\eeas
for every $p>1$ under Conditions $[M1]$ and $[M2]$. 
Therefore, Condition $[L4]$(iv) for $\beta_1=1/2$ was verified. 
It is also possible to show $[L4]$(iii) in a similar fashion by using the mixing property and Sobolev's inequality. 
Condition $[L4]$(i) is already checked in (\ref{201912010230}). 
Thus, Condition $[L4]$ has been verified.  
\begin{en-text}
Therefore, 
\beas 
\Gamma_T^i(\rho^i)
\yeq
-T^{-1}\partial_{\rho^i}^2\bbH_T^{(i)}(\rho^i)
&\to^p&
E\bigg[{\sf V}_0(\bbY^i(0),\rho^i)\otimes(\bbY(0))^{\otimes2}\Lambda(\lambda_0(0),\bbX(0))r^i(0,\theta^*)\bigg]
=:\Gamma^i(\rho^i)
\eeas
as $T\to\infty$ uniformly in $\rho^i$. 
\end{en-text}
\begin{en-text}
Condition $[L4]$(i) is also proved in a similar manner for 
\beas 
\Delta_T &=& T^{-1/2}\partial_{(\theta,\rho)}\bbH_T(\theta^*,\rho^*)
\eeas
by using (\ref{201911301926}) and (\ref{201911301925}). 
\end{en-text}

\subsection{Conditions $[L2]$ and $[L3]$}
We see 
\beas 
\partial_{(\theta,\rho)}^2\bbY(\theta,\rho)=\Gamma(\theta,\rho),
\eeas 
and by $[M3]$, we conclude $\bbY(\theta,\rho)$ is strictly convex function on 
$\overline{\Theta}\times\overline{\calr}=\overline{\Theta}\times\Pi_{i\in\bbI}\overline{\calr}_i$. 
For some neighborhood $U$ of $(\theta^*,\rho^*)$ and some positive number $\chi_1$, 
\beas 
\bbY(\theta,\rho)\leq -\chi_1|(\theta,\rho)-(\theta^*,\rho^*)|^2\qquad \big((\theta,\rho)\in U\big)
\eeas
by the non-degeneracy of $\Gamma(\theta^*,\rho^*)$. 
Moreover, 
$\sup_{(\theta,\rho)\in(\Theta\times\calr)\setminus U}\bbY(\theta,\rho)<0$. 
In fact, if there was a point $(\theta^+,\rho^+)\not\in U$ 
such that $\bbY(\theta^+,\rho^+)=0$, then at a point on the segment connecting 
$(\theta^*,\rho^*)$ and $(\theta^+,\rho^+)$, $\Gamma(\theta,\rho)$ would degenerate, 
and this contradicts $[M3]$. 
As a consequence, Condition $[L2]$ is verified for $\rho=2$ and some (deterministic) positive number $\chi_0$ 
since the parameter space is bounded. 
Condition $[L3]$ is now obvious. 

\subsection{Proof of Theorem \ref{201911220134}}
We have verified Conditions $[L1]$-$[L4]$ in the present situation. 
Theorem \ref{201911220134} now follows from Theorems \ref{201911221015} and \ref{201912011335}. 
\qed\halflineskip

\section{Quasi-likelihood analysis}\label{201911221017}
\label{sec:QLArefresher}
This section recalls the quasi-likelihood analysis. 
Let $\Theta$ be a bounded open set in $\bbR^\sfp$. 
Given a probability space $(\Omega,\calf,P)$, 
suppose that $\bbH_T:\Omega\times\overline{\Theta}\to\bbR$ is of class $C^3$, 
that is, the mapping 
$\Theta\ni\theta\mapsto\bbH_T(\omega,\theta)\in\bbR^\sfp$ is continuously extended to $\overline{\Theta}$ 
and of class $C^3$
for every $\omega\in\Omega$, and the mapping 
$\Omega\ni\omega\mapsto\bbH_T(\omega,\theta)\in\bbR^\sfp$ is measurable 
for every $\theta\in\Theta$. 
Let $\Gamma$ be a $\sfp\times\sfp$ random matrix. 

Let $\theta^*\in\Theta$. 
For a sequence $a_T\in GL(\sfp)$ satisfying $\lim_{T\to\infty}|a_T|=0$, let 
\bea\label{201912010242} 
\Delta_T[u] = \partial_\theta\bbH_T(\theta^*)a_T
\quad\text{and}\quad
\Gamma_T(\theta) =
-a_T^\star\partial_\theta^2\bbH_T(\theta)a_T
\eea
where $\star$ denotes the matrix transpose. 
We consider a random field 
\bea\label{201912010222}
\bbY_T(\theta) = b_T^{-1}\big(\bbH_T(\theta)-\bbH_T(\theta^*)\big),
\eea 
which will be assumed to converge to a random field
$\bbY:\Omega\times\Theta\to\bbR$. 
Only for simplifying presentation, 
we will assume that $a_T=b_T^{-1/2}I_\sfp$ for diverging sequence $(b_T)_{T>0}$of positive numbers, 
where $I_\sfp$ is the identity matrix. 
In what follows, we fix a positive number $L$.

We will give a simplified exposition of \cite{Yoshida2011} on the polynomial type large deviation inequality. 
Let $\alpha$, $\beta_1$, $\beta_2$, $\rho$, $\rho_1$ and $\rho_2$ be numbers. 

\bd
\im[{\bf[$L1$]}] 
The numbers $\alpha$, $\beta_1$, $\beta_2$, $\rho$, 
$\rho_1$ and $\rho_2$ satisfy the following inequalities: 
\beas 
&&0<\alpha<1,\quad
0<\beta_1<1/2,\quad 0<\rho_1<\min\{1,\alpha(1-\alpha)^{-1},2\beta_1(1-\alpha)^{-1}\},\\
&&\alpha\rho<\rho_2,\quad\beta_2\geq0\quad\text{and}\quad
1-2\beta_2-\rho_2>0.
\eeas

Let $\beta=\alpha(1-\alpha)^{-1}$.

\im[{\bf[$L2$]}] 
There is a positive random variable $\chi_0$ such that 
\beas 
\bbY(\theta)\>=\>\bbY(\theta)-\bbY(\theta^*)
&\leq& -\chi_0|\theta-\theta^*|^\rho
\eeas
for all $\theta\in\Theta$. 

\im[{\bf[$L3$]}] 
There exists a $C_L$ such that
\beas 
P\big[\chi_0\leq r^{-(\rho_2-\alpha\rho)}\big] &\leq& \frac{C_L}{r^L}\quad(r>0)
\eeas
and 
\beas 
P\big[\lambda_{\text{min}}(\Gamma)<4r^{-\rho_1}\big]&\leq&\frac{C_L}{r^L}\quad(r>0).
\eeas

\im[{\bf[$L4$]}] 
{\bf (i)} 
For $M_1=L(1-\rho_1)^{-1}$, 
$\displaystyle
\sup_{T>0}E\big[|\Delta_T|^{M_1}\big] <\infty.
$
\bd
\im[(ii)] For $M_2=L(1-2\beta_2-\rho_2)^{-1}$, 
\beas 
\sup_{T>0}E\bigg[\bigg(\sup_{h:|h|\geq b_T^{-\alpha/2}}
b_T^{\half-\beta_2}\big|\bbY_T(\theta^*+h)-\bbY(\theta^*+h)\big|\bigg)^{M_2}\bigg]
&<&\infty. 
\eeas
\im[(iii)] For $M_3=L(\beta-\rho_1)^{-1}$, 
\beas 
\sup_{T>0}E\bigg[\bigg(b_T^{-1}\sup_{\theta\in\Theta}\big|\partial_\theta^3\bbH_T(\theta)
\big|\bigg)^{M_3}\bigg] &<& \infty. 
\eeas
\im[(iv)] 
For $M_4=L\big(2\beta_1(1-\alpha)^{-1}-\rho_1\big)^{-1}$, 
\beas 
\sup_{T>0}E\bigg[\bigg(b_T^{\beta_1}
\big|\Gamma_T(\theta^*)-\Gamma\big|\bigg)^{M_4}\bigg] 
&<& \infty.
\eeas
\ed
\ed

Let $\bbU_T=\{u\in\bbR^\sfp;\>\theta^*+a_Tu\in\Theta\}$ and 
$\bbV_T(r)=\{u\in\bbU_T;\>|u|\geq r\}$ for $r>0$.

\begin{theorem}\label{201911220753}{\rm (\cite{Yoshida2011})} 
Suppose that Conditions $[L1]$-$[L4]$ are satisfied. Then 
there exists a constant $C$ such that 
\beas 
P\bigg[\sup_{u\in\bbV_T(r)}\bbZ_T(u)\geq \exp\big(-2^{-1}r^{2-(\rho_1\vee\rho_2)}\big)
\bigg]&\leq&\frac{C}{r^L}
\eeas
for all $T>0$ and $r>0$. 
Here the supremum of the empty set should read $-\infty$ by convention. 
\end{theorem}

We comments some points. 
Parameters satisfying $[L1]$ exist. 
Nondegeneracy conditions in $[\cala3]$ are obvious in ergodic cases. 
In this paper, we will apply Theorem \ref{201911220753} under ergodicity of the stochastic system. 
Theorem \ref{201911220753} asserts a polynomial type large deviation inequality 
can be obtained once the boundedness of moments of some random variables is verified. 
Condition $[L4]$ is easy to obtain because each variable is usually a simple additive functional. 
The polynomial type large deviation inequality in Theorem \ref{201911220753} 
enables us to easily apply the scheme by \cite{IbragimovHascprimeminskiui1981} and \cite{kutoyants1984parameter,kutoyants2012statistical}  
to various dependence structures.

Let $u\in\bbR^\sfp$. 
Define $r_T(u)$ $(u\in\bbU_T)$ by 
\bea\label{201912011356} 
\bbZ_T(u) &=& \exp\bigg(\Delta_T[u]-\half\Gamma[u^{\otimes2}]+r_T(u)\bigg)
\quad(u\in\bbU_T)
\eea
It is said that 
$\bbZ_T$ is locally asymptotically quadratic (LAQ) at $\theta^*$ 
if $r_T(u)\to^p0$ as $T\to\infty$ for every $u\in\bbR^\sfp$, and hence 
$\log\bbZ_T(u)$ is asymptotically approximated by a random quadratic function of $u$.

We will confine our attention to a very standard case where 
$\bbZ_T$ is locally asymptotically mixed normal, though the general theory of the quasi-likelihood analysis 
is framed more generally.  

Any measurable mapping $\hat{\theta}_T^M:\Omega\to\overline{\Theta}$ is called 
a quasi-maximum likelihood estimator (QMLE) for $\bbH_T$ if 
\beas 
\bbH_T(\hat{\theta}_T^M) &=& \max_{\theta\in\overline{\Theta}}\bbH_T(\theta).
\eeas
When $\bbH_T$ is continuous on the compact $\overline{\Theta}$, such a measurable function always exists, 
which is ensured by the measurable selection theorem. 
Let $\hat{u}_T^M=a_T^{-1}(\hat{\theta}_T^M-\theta^*)$ for the QMLE $\hat{\theta}_T^M$. 

\begin{theorem}\label{201911221015}
Let $L>\sfp>0$. 
Suppose that Conditions $[L1]$-$[L4]$ are satisfied and that 
$(\Delta_T,\Gamma)\to^d(\Gamma^{1/2}\zeta,\Gamma)$ as $T\to\infty$, where 
$\zeta$ is a $\sfp$-dimensional standard Gaussian random vector independent of $\Gamma$. 
Then 
\beas 
E\big[f(\hat{u}_T^M)\big] &\to& \bbE\big[f(\hat{u})\big]\quad(T\to\infty)
\eeas
for $\hat{u}=\Gamma^{-1/2}\zeta$ and for 
any $f\in C(\bbR^\sfp)$ satisfying $\lim_{|u|\to\infty}|u|^{-p}|f(u)|<\infty$. 
\end{theorem}
\proof 
We will sketch the proof to convey the concepts of the quasi-likelihood analysis to the reader. 
See \cite{Yoshida2011} for details. 
The space $\hat{C}(\bbR^\sfp)$ is the linear space of all continuous functions $f:\bbR^\sfp\to\R$ 
satisfying $\lim_{|u|\to\infty}f(u)=0$. The space $\hat{C}(\bbR^\sfp)$ becomes a separable Banach space 
equipped with the supremum norm $\|f\|_\infty=\sup_{u\in\bbR^\sfp}|f(u)|$. 
Moreover, $\hat{C}(\bbR^\sfp)$ is regarded as a measurable space with the Borel $\sigma$-field. 
Let 
\bea\label{201912011515} 
\bbZ(u) &=& \exp\bigg(\Gamma^{1/2}\zeta[u]-\half\Gamma[u^{\otimes2}]\bigg)
\eea
for $u\in\bbR^\sfp$. 

The term $r_T(u)$ admits the expression 
\bea\label{26771120-4} 
r_T(u) &=& 
\int_0^1(1-s)\big\{\Gamma[u^{\otimes2}]-\Gamma_T(\theta^*+sa_Tu)[u^{\otimes2}]\big\}ds
\eea
for $u$ such that $|u|\leq b_T^{(1-\alpha)/2}$ 
and 
$T$ such that $B(\theta^*,b_T^{-\alpha/2})\subset\Theta$. 
In this situation, we can apply Taylor's formula even though 
the whole $\Theta$ is not convex. 
Condition $[L4]$ (iii) and the convergence of $\Delta_T$ ensures tightness of 
the random fields $\big\{\bbZ_T|_{\overline{B(0,R)}}\big\}_{T>T_0}$ for every $R>0$, where 
$B(0,R)=\{u\in\bbR^\sfp\}$ and $T_0$ is a sufficiently large number depending on $R$. 
Combining this property with the polynomial type large deviation inequality given by Theorem \ref{201911220753}, 
we obtain the convergence $\bbZ_T\to\bbZ$ in $\hat{C}(\bbR^\sfp)$ for the random field $\bbZ_T$ 
extended as an element of $\hat{C}(\bbR^\sfp)$ so that 
$\sup_{\bbR^\sfp\setminus\bbU_T}\bbZ_T(u)\leq\sup_{u\in\partial\bbU_T}\bbZ_T(u)$. 
Consequently, $\hat{u}_T\to\hat{u}=\text{argmax}_{u\in\bbR^\sfp}\bbZ(u)$. 
It is known that a measurable version of extension of $\bbZ_T$ exists.

\begin{en-text}
In the proof of Theorem \ref{201911220753}, it is proved that 
$\bbZ_T$ is {\bf locally asymptotically quadratic} (LAQ) at $\theta^*$, 
that is, $r_T(u)\to^p0$ as $T\to\infty$, and hence 
$\log\bbZ_T(u)$ is asymptotically approximated by a random quadratic function of $u$. 
More precisely, 
\begin{proposition}\label{26771120-6}
Assume $[L1]$ and $[L4]$ (iii) and (iv). Then 
there exists a constant $C_L$ such that 
\beas 
\sup_{T>0}P\bigg[
\sup_{u\in\bbU_T(r)}(1+|u|^2)^{-1}\big|r_T(u)\big|\geq r^{-\rho_1}\bigg]
&\leq& 
\frac{C_L}{r^L}\quad(r>0).
\eeas
\end{proposition}
A special case of Proposition \ref{26771120-6} is $\Gamma$ is the case where 
a deterministic positive definite matrix and $\Delta_T\to N_\sfp(0,\Gamma)$. 
Then the finite-dimensional convergence $\bbZ_T\to^{d_f}\bbZ$ holds for 
the random field (defined on some probability space)  
$\bbZ(u)=\exp\big(\Gamma^{1/2}\zeta-2^{-1}\Gamma[u^{\otimes2}]\big)$ where $\zeta$ is 
a $\sfp$-dimensional standard normal random vector. 
\end{en-text}
A polynomial type large deviation, even weaker than the one in Theorem \ref{201911220753}, 
serves to show $L^q$-boundedness of $\{|\hat{u}_T|^q\}$ for $L>q>p$. 
Then the family $\{\hat{u}_T\}$ is uniformly integrable, and hence we obtain the convergence of 
$E[f(\hat{u}_T)]$. 
\begin{en-text}
\begin{theorem}\label{281016-3}
Let $L>p>0$. Suppose that there exists a constant $C$ such that 
\beas 
P\bigg[\sup_{u\in\bbV_T(r)}\bbZ_T(u)\geq1\bigg] &\leq& \frac{C}{r^L}
\eeas
for all $T>0$ and $r>0$. 
Then any sequence of QMLE's $\hat{\theta}_T$ admits 
\beas 
\sup_{T>0}E\big[|\hat{u}_T|^p\big] &<& \infty. 
\eeas
\end{theorem}
\end{en-text}
\qed\halflineskip

\begin{remark}\rm 
In Theorem \ref{201911221015}, if 
$\Delta_T\to^d\Gamma^{1/2}\zeta$ $\calf$-stably, 
then $(\Delta_T,\Gamma)\to^d(\Gamma^{1/2}\zeta,\Gamma)$ and 
$\hat{u}^M_T\to\hat{u}$ $\calf$-stably. 
\end{remark}

An advantage of the quasi-likelihood analysis is that the asymptotic behavior of the quasi-Bayesian estimator 
can be obtained as well as that of the quasi-maximum likelihood estimator and its moments convergence. 
The mapping 
\beas 
\hat{\theta}_T^B 
&=&
\bigg[\int_\Theta\exp\big(\bbH_T(\theta)\big)\varpi(\theta)d\theta\bigg]^{-1}
\int_\Theta\theta\exp\big(\bbH_T(\theta)\big)\varpi(\theta)d\theta
\eeas
is called a quasi-Bayesian estimator (QBE) with respect to the prior density $\varpi$. 
The QBE $\hat{\theta}_T^B$ takes values in the convex-hull of $\overline{\Theta}$. 
We will assume $\varpi$ is continuous and 
$0<\inf_{\theta\in\Theta}\varpi(\theta)\leq\sup_{\theta\in\Theta}\varpi(\theta)<\infty$. 
We will give a concise exposition in the following among many possible ways. 
The reader is referred to \cite{Yoshida2011} for further information.  
Recall that $\sfp$ is the dimension of $\Theta$, and $B(R)$ denotes the open ball of radius $R$ centered at the origin. 
$C(\overline{B(R)})$ is the space of all continuous functions on $\overline{B(R)}$, and 
it is equipped with the supremum norm. 
Let $\bbV_T(r)=\{u\in\bbU_T;\>|u|\geq r\}$. 
As before, $\hat{u}=\Gamma^{-1/2}\zeta$ with a $\sfp$-dimensional standard Gaussian random vector $\zeta$ 
independent of $\Gamma$. 
Write $\hat{u}_T^B=a_T^{-1}(\hat{\theta}_T^B-\theta^*)$. 
\begin{theorem}\label{201912011301}
Let $p\geq1$, $L>p+1$, $D>\sfp+p$. 
Suppose that 
$(\Delta_T,\Gamma)\to^d(\Gamma^{1/2}\zeta,\Gamma)$ as $T\to\infty$, where 
$\zeta$ is a $\sfp$-dimensional standard Gaussian random vector independent of $\Gamma$. 
Moreover, suppose the following conditions are satisfied. 
\bd\im[(i)] 
For every $R>0$, 
\bea\label{201912011341}
\bbZ_T|_{\overline{B(R)}}&\to&^d\bbZ|_{\overline{B(R)}}
\quad\text{in }C(\overline{B(R)})
\eea
as $T\to\infty$, where $\bbZ$ is given in (\ref{201912011515}). 
\im[(ii)] There exist positive constants $T_0$, $C_1$ and $C_2$ such that 
\bea\label{201912011344} 
P\bigg[\sup_{\bbV_T(r)}\bbZ_T\geq C_1r^{-D}\bigg]&\leq& C_2r^{-L}
\eea
for all $T\geq T_0$ and $r>0$. 
\im[(iii)] For some $T_0>0$, 
\bea\label{201912011345}
\sup_{T\geq T_0}E\bigg[\bigg(\int_{\bbU_T}\bbZ_T(u)\bigg)^{-1}\bigg] &<& \infty.
\eea
\ed
Then 
\bea\label{201912011334} 
E\big[f(\hat{u}_T^B)\big] &\to& E\big[f(\hat{u})\big]
\eea
as $T\to\infty$ for any continuous function $f:\bbR^sfp\to\bbR$ satisfying 
$\sup_{u\in\bbR^\sfp}\big\{(1+|u|)^{-p}|f(u)|\big\}<\infty$. 
\end{theorem}
\proof We will give a brief summary of the proof; see \cite{Yoshida2011} for details. 
The variable $\hat{u}_T^B$ has the expression 
\beas 
\hat{u}_T^B
&=& 
\bigg[\int_{\bbU_T}\bbZ_T(u)\varpi(\theta^*+a_Tu)du\bigg]^{-1}
\int_{\bbU_T} u\bbZ_T(u)\varpi(\theta^*+a_Tu)du
\eeas
By (\ref{201912011344}) and the properties of $\varpi$, 
we can approximate $\hat{u}_T^B$ by 
\beas 
\tilde{u}_T &=& \bigg[\int_{B(R)}\bbZ_T(u)du\bigg]^{-1}
\int_{B(R)} u\bbZ_T(u)du
\eeas
for paying small error when $R$ is large. 
By (\ref{201912011341}), 
\beas 
\tilde{u}_T &\to^d& 
\bigg[\int_{B(R)}\bbZ(u)du\bigg]^{-1}
\int_{B(R)} u\bbZ(u)du=:\hat{u}(R). 
\eeas
The random field $\bbZ$ inherits a tail estimate from (\ref{201912011344}), and hence 
$\hat{u}(R)$ is approximated by 
\beas 
\bigg[\int_{\bbR^\sfp}\bbZ(u)du\bigg]^{-1}\int_{\bbR^\sfp} u\bbZ(u)du
\yeq \Gamma^{-1/2}\zeta\yeq \hat{u}.
\eeas
Combining these estimates, we can conclude $\hat{u}_T^B\to^d\hat{u}$ as $T\to\infty$. 
Convergence of the expectation is a consequence of uniform integrability of $|\hat{u}_T^B|^p$ ensured by 
(\ref{201912011344}). 
\qed\halflineskip
\begin{remark}\rm 
(a) It is possible to relax the conditions of Theorem \ref{201912011301} to only ensure 
the convergence $\hat{u}^B_T\to\hat{u}$. 
(b) In Theorem \ref{201912011301}, if $(\Delta_T,\Gamma)\to^d(\Gamma^{1/2}\zeta,\Gamma)$ and 
$\hat{u}^B_T\to\hat{u}$ $\calf$-stably. 
(c) Usually, the condition (iii) of Theorem \ref{201912011301} is easily verified; See Lemma 2 of \cite{Yoshida2011}. 
\end{remark}
The following result follows from Theorem \ref{201912011301}. 
\begin{theorem}\label{201912011335}
Let $p>\sfp$ and 
\beas 
L>\max\bigg\{p+1,\sfp(\beta-\rho_1),\sfp(2\beta_1(1-\alpha)^{-1}-\rho_1)\bigg\}. 
\eeas
Suppose that Conditions $[L1]$-$[L4]$ are satisfied and that $E[|\Gamma|^p]<\infty$. 
$(\Delta_T,\Gamma)\to^d(\Gamma^{1/2}\zeta,\Gamma)$ as $T\to\infty$, where 
$\zeta$ is a $\sfp$-dimensional standard Gaussian random vector independent of $\Gamma$. 
Then 
\beas 
E\big[f(\hat{u}_T^B)\big] &\to& \bbE\big[f(\hat{u})\big]\quad(T\to\infty)
\eeas
for $\hat{u}=\Gamma^{-1/2}\zeta$ and for 
any $f\in C(\bbR^\sfp)$ satisfying $\lim_{|u|\to\infty}|u|^{-p}|f(u)|<\infty$. 
\end{theorem}
\proof 
The convergence (\ref{201912011341}) holds, as shown in the proof of Theorem \ref{201911221015}. 
The polynomial type large deviation inequality (\ref{201912011344}) is a consequence of 
Theorem \ref{201911220753}; the number $D$ is arbitrary. 
Fix $\delta>0$. Then there exists $T_0>0$ such that $B(\delta)\subset\Theta$. 
In particular,  
$r_T(u)$ admits the representation (\ref{26771120-4}) for all $u\in B(\delta)$. 
Since $M_3=L(\beta-\rho_1)^{-1}>\sfp$, 
$M_4=L(2\beta_1(1-\alpha)^{-1}-\rho_1)^{-1}>\sfp$ and $p>\sfp$, we have 
$p':=\min\{M_3,M_4,p\}>\sfp$ and 
\beas 
E[|r_T(u)|^{p'}] &\leq& C_0|u|^{p'}\quad(u\in B(\delta))
\eeas
for some constant $C_0$. 
Then Lemma 2 of \cite{Yoshida2011} gives the estimate 
\beas 
E\bigg[\bigg(\int_{B(\delta)}\bbZ_T(u)du\bigg)^{-1}\bigg] &\leq& C_1
\eeas
by a constant $C_1$ depending on $(p',\sfp,\delta,C_0)$ and the supremums appearing in $[L4]$(i),(iii),(iv), 
but $C_1$ is independent of $T\geq T_0$. 
Therefore (\ref{201912011345}) holds true. 
Thus, we can apply Theorem \ref{201912011301} to conclude the proof. 
\qed\halflineskip

\section{List of stocks}
\label{sec:ricList}
Table \ref{table:ricList} lists all the stocks investigated in the paper. For each stock, the total number of days available in the sample is given. Note that for lack of usage time allotment on the computational resources used for this paper, some trading days for  few very liquid stocks were not used for some of the marked ratio models tested in Section \ref{subsec:InSampleQAIC}. In this case, only the trading days where \emph{all} models have been computed have been used. This is the last column of the table.

\begin{table}
\begin{center}
\footnotesize
\begin{tabular}{ccccc}
\hline
\multirow{2}{*}{RIC} & \multirow{2}{*}{Company} & \multirow{2}{*}{Sector} & Number of trading & Number of trading
\\ & & & days in sample & days used in QAIC \\ \hline
AIRP.PA & Air Liquide & Healthcare / Energy & 238 & 238 \\
BNPP.PA & BNP Paribas & Banking & 224 & 62 \\ 
EDF.PA & Electricite de France & Energy & 236 & 236 \\ 
LAGA.PA & Lagardère & Media & 142 & 142 \\ 
CARR.PA & Carrefour & Retail & 229 & 229\\ 
BOUY.PA & Bouygues & Construction / Telecom & 228 & 228 \\ 
ALSO.PA & Alstom & Transport & 229 & 229 \\
ACCP.PA & Accor & Hotels & 227 & 227 \\ 
ALUA.PA & Alcatel & Networks / Telecom & 234 & 234 \\ 
AXAF.PA & Axa & Insurance & 236 & 131 \\ 
CAGR.PA & Crédit Agricole & Banking & 235 & 235 \\ 
CAPP.PA & Cap Gemini & Technology Consulting & 232 & 232 \\ 
DANO.PA & Danone & Food & 229 & 229 \\ 
ESSI.PA & Essilor & Optics & 228 & 228 \\ 
LOIM.PA & Klepierre & Finance & 221 & 221 \\ 
LVMH.PA & Louis Vuitton Moët Hennessy & Luxury & 233 & 198 \\ 
MICP.PA & Michelin & Tires & 229 & 229 \\ 
OREP.PA & L'Oréal & Cosmetics & 233 & 233 \\ 
PERP.PA & Pernod Ricard & Spirits & 224 & 224 \\ 
PEUP.PA & Peugeot & Automotive & 151 & 151 \\ 
PRTP.PA & Kering & Luxury & 227 & 227 \\ 
PUBP.PA & Publicis & Communication & 223 & 223 \\ 
RENA.PA & Renault & Automotive & 228 & 172 \\ 
SAF.PA & Safran & Aerospace / Defense & 232 & 232 \\ 
TECF.PA & Technip & Energy & 225 & 225 \\ 
TOTF.PA & Total & Energy & 232 & 75 \\ 
VIE.PA & Veolia & Energy / Environment & 234 & 234 \\ 
VIV.PA & Vivendi & Media & 234 & 234 \\ 
VLLP.PA & Vallourec & Materials & 228 & 228 \\ 
VLOF.PA & Valeo & Automotive & 221 & 212 \\ 
SASY.PA & Sanofi & Healthcare & 229 & 97 \\ 
SCHN.PA & Schneider Electric & Energy & 224 & 164 \\ 
SGEF.PA & Vinci & Construction & 229 & 229 \\ 
SGOB.PA & Saint Gobain & Materials & 234 & 180 \\ 
SOGN.PA & Société Générale & Banking & 229 & 103 \\ 
STM.PA & ST Microelectronics & Semiconductor & 227 & 227 \\ \hline
\normalsize
\end{tabular}
\caption{List of stocks investigated in this paper. Sample consists of the whole year 2015, representing roughly 230 trading days for all stocks except LAGA.PA and PEUP.PA which are missing roughly 70 trading days.}
\label{table:ricList}
\end{center}
\end{table}

\section{In-sample AIC selection - Detailed results}
\label{sec:AICDetailedResults}
\begin{table}
\centering
\tiny
\input{AIC-sideModelRanking.tex}
\caption{Side determination - AIC most selected models by stock (covariates on the first line, frequency on the second line)}
\label{table:AIC-side-ModelRanking}
\end{table}
\begin{table}
\centering
\tiny
\input{AIC-aggBModelRanking.tex}
\caption{Bid aggressiveness determination - AIC most selected models by stock (covariates on the first line, frequency on the second line)}
\label{table:AIC-aggB-ModelRanking}
\end{table}
\begin{table}
\centering
\tiny
\input{AIC-aggAModelRanking.tex}
\caption{Ask aggressiveness determination - AIC most selected models by stock (covariates on the first line, frequency on the second line)}
\label{table:AIC-aggA-ModelRanking}
\end{table}

\end{document}

%% file: nakamacro-markedVersion.tex
\def\infm{{\infty\text{--}}}
\def\inftym{\infm}
\def\koko{{\coloroy{koko}}}
\def\bd{\begin{description}}
\def\ed{\end{description}}

\def\D2{\bbD_{2,\infty-}}

\def\D{{\bf D}}

\def\R{{\bf R}}

\def\cala{{\cal A}}
\def\calb{{\cal B}}

\def\calf{{\cal F}}

\def\calj{{\cal J}}

\def\calr{{\cal R}}

\def\ds{\displaystyle}
\def\yeq{\>=\>}

\def\sfp{{\sf p}}

\def\simleq{\ \raisebox{-.7ex}{$\stackrel{{\textstyle <}}{\sim}$}\ }

\def\half{\frac{1}{2}}

\def\halflineskip{\vspace*{3mm}}
\def\nn{\nonumber}
\def\be{\begin{equation}}
\def\ee{\end{equation}}
\def\bea{\begin{eqnarray}}
\def\eea{\end{eqnarray}}
\def\beas{\begin{eqnarray*}}
\def\eeas{\end{eqnarray*}}
\def\bi{\begin{itemize}}
\def\ei{\end{itemize}}
\def\im{\item}
\def\bd{\begin{description}}
\def\ed{\end{description}}
\newcommand{\bbD}{{\mathbb D}}
\newcommand{\bbE}{{\mathbb E}}
\newcommand{\bbF}{{\mathbb F}}

\newcommand{\bbH}{{\mathbb H}}
\newcommand{\bbI}{{\mathbb I}}
\newcommand{\bbJ}{{\mathbb J}}
\newcommand{\bbK}{{\mathbb K}}

\newcommand{\bbN}{{\mathbb N}}

\newcommand{\bbR}{{\mathbb R}}

\newcommand{\bbU}{{\mathbb U}}
\newcommand{\bbV}{{\mathbb V}}

\newcommand{\bbX}{{\mathbb X}}
\newcommand{\bbY}{{\mathbb Y}}
\newcommand{\bbZ}{{\mathbb Z}}

%% file: AIC-sideModelRanking.tex
\begin{tabular}{rcccc}
\toprule
{} &        0 &        1 &        2 &          3 \\
\midrule
ACCP.PA &  1234689 &    14689 &   124689 &  123456789 \\
 &     0.34 &     0.25 &     0.23 &       0.10 \\
AIRP.PA &    14689 &   124689 &  1234689 &    1456789 \\
 &     0.28 &     0.26 &     0.21 &       0.11 \\
ALSO.PA &   124689 &  1234689 &    14689 &  123456789 \\
 &     0.28 &     0.25 &     0.23 &       0.12 \\
ALUA.PA &  1234689 &   124689 &    14689 &  123456789 \\
 &     0.32 &     0.21 &     0.18 &       0.14 \\
AXAF.PA &    14689 &  1234689 &   124689 &    1456789 \\
 &     0.28 &     0.27 &     0.21 &       0.12 \\
BNPP.PA &  1234689 &   124689 &    14689 &   12456789 \\
 &     0.32 &     0.27 &     0.19 &       0.10 \\
BOUY.PA &  1234689 &    14689 &   124689 &  123456789 \\
 &     0.30 &     0.27 &     0.22 &       0.09 \\
CAGR.PA &    14689 &  1234689 &   124689 &    1456789 \\
 &     0.34 &     0.24 &     0.20 &       0.10 \\
CAPP.PA &  1234689 &   124689 &    14689 &  123456789 \\
 &     0.31 &     0.25 &     0.22 &       0.09 \\
CARR.PA &  1234689 &    14689 &   124689 &    1456789 \\
 &     0.32 &     0.28 &     0.23 &       0.07 \\
DANO.PA &    14689 &  1234689 &   124689 &  123456789 \\
 &     0.29 &     0.27 &     0.23 &       0.10 \\
EDF.PA  &  1234689 &    14689 &   124689 &  123456789 \\
  &     0.30 &     0.30 &     0.22 &       0.08 \\
ESSI.PA &    14689 &  1234689 &   124689 &    1456789 \\
 &     0.30 &     0.29 &     0.20 &       0.08 \\
LAGA.PA &    14689 &   124689 &  1234689 &    1456789 \\
 &     0.46 &     0.18 &     0.16 &       0.08 \\
LOIM.PA &  1234689 &   124689 &    14689 &  123456789 \\
 &     0.29 &     0.27 &     0.19 &       0.11 \\
LVMH.PA &    14689 &   124689 &  1234689 &    1456789 \\
 &     0.29 &     0.25 &     0.21 &       0.09 \\
MICP.PA &  1234689 &    14689 &   124689 &  123456789 \\
 &     0.32 &     0.27 &     0.23 &       0.08 \\
OREP.PA &    14689 &  1234689 &   124689 &    1456789 \\
 &     0.31 &     0.24 &     0.24 &       0.11 \\
PERP.PA &    14689 &  1234689 &   124689 &  123456789 \\
 &     0.30 &     0.24 &     0.22 &       0.06 \\
PEUP.PA &   124689 &  1234689 &    14689 &    1456789 \\
 &     0.30 &     0.28 &     0.22 &       0.07 \\
PRTP.PA &  1234689 &    14689 &   124689 &    1456789 \\
 &     0.30 &     0.27 &     0.22 &       0.06 \\
PUBP.PA &  1234689 &   124689 &    14689 &  123456789 \\
 &     0.34 &     0.26 &     0.23 &       0.06 \\
RENA.PA &  1234689 &   124689 &    14689 &   12456789 \\
 &     0.34 &     0.25 &     0.23 &       0.06 \\
SAF.PA  &  1234689 &    14689 &   124689 &    1456789 \\
  &     0.32 &     0.24 &     0.24 &       0.09 \\
SASY.PA &    14689 &   124689 &  1234689 &    1456789 \\
 &     0.30 &     0.29 &     0.27 &       0.05 \\
SCHN.PA &    14689 &  1234689 &   124689 &    1456789 \\
 &     0.29 &     0.27 &     0.25 &       0.09 \\
SGEF.PA &  1234689 &   124689 &    14689 &  123456789 \\
 &     0.33 &     0.27 &     0.21 &       0.10 \\
SGOB.PA &  1234689 &    14689 &   124689 &  123456789 \\
 &     0.32 &     0.29 &     0.22 &       0.06 \\
SOGN.PA &  1234689 &    14689 &   124689 &    1456789 \\
 &     0.36 &     0.30 &     0.17 &       0.08 \\
STM.PA  &    14689 &  1234689 &   124689 &   12456789 \\
  &     0.32 &     0.25 &     0.22 &       0.07 \\
TECF.PA &  1234689 &    14689 &   124689 &  123456789 \\
 &     0.33 &     0.22 &     0.22 &       0.12 \\
TOTF.PA &  1234689 &   124689 &    14689 &  123456789 \\
 &     0.33 &     0.28 &     0.16 &       0.12 \\
VIE.PA  &    14689 &  1234689 &   124689 &    1456789 \\
  &     0.33 &     0.26 &     0.22 &       0.08 \\
VIV.PA  &    14689 &  1234689 &   124689 &    1456789 \\
  &     0.28 &     0.25 &     0.24 &       0.09 \\
VLLP.PA &  1234689 &   124689 &    14689 &  123456789 \\
 &     0.38 &     0.27 &     0.13 &       0.12 \\
VLOF.PA &    14689 &  1234689 &   124689 &    1456789 \\
 &     0.32 &     0.32 &     0.16 &       0.09 \\
\bottomrule
\end{tabular}

%% file: AIC-aggBModelRanking.tex
\begin{tabular}{rcccc}
\toprule
{} &      0 &      1 &       2 &        3 \\
\midrule
ACCP.PA &  14689 &    146 &     189 &    14567 \\
 &   0.18 &   0.17 &    0.09 &     0.09 \\
AIRP.PA &   1458 &  14689 &   14567 &      146 \\
 &   0.32 &   0.16 &    0.13 &     0.11 \\
ALSO.PA &    146 &    189 &   14689 &    14567 \\
 &   0.23 &   0.14 &    0.10 &     0.10 \\
ALUA.PA &    189 &    146 &   14689 &    14567 \\
 &   0.19 &   0.16 &    0.15 &     0.11 \\
AXAF.PA &   1458 &  14689 &   14567 &     1246 \\
 &   0.33 &   0.21 &    0.15 &     0.06 \\
BNPP.PA &  14689 &  14567 &  124689 &     1458 \\
 &   0.26 &   0.21 &    0.11 &     0.06 \\
BOUY.PA &    146 &  14689 &     189 &     1458 \\
 &   0.19 &   0.16 &    0.13 &     0.11 \\
CAGR.PA &   1458 &    146 &   14689 &    14567 \\
 &   0.23 &   0.18 &    0.17 &     0.14 \\
CAPP.PA &    146 &  14689 &     189 &    14567 \\
 &   0.25 &   0.17 &    0.10 &     0.10 \\
CARR.PA &  14689 &    146 &    1458 &    14567 \\
 &   0.21 &   0.19 &    0.17 &     0.12 \\
DANO.PA &  14689 &    146 &    1458 &    14567 \\
 &   0.21 &   0.17 &    0.16 &     0.12 \\
EDF.PA  &    146 &  14689 &    1458 &    14567 \\
  &   0.24 &   0.17 &    0.10 &     0.09 \\
ESSI.PA &    146 &  14567 &    1458 &    14689 \\
 &   0.17 &   0.17 &    0.17 &     0.16 \\
LAGA.PA &    146 &    189 &   14689 &     1679 \\
 &   0.33 &   0.15 &    0.09 &     0.09 \\
LOIM.PA &    146 &    189 &   14689 &    14567 \\
 &   0.19 &   0.19 &    0.13 &     0.09 \\
LVMH.PA &   1458 &  14567 &   14689 &      146 \\
 &   0.27 &   0.25 &    0.18 &     0.08 \\
MICP.PA &    146 &  14689 &   14567 &      189 \\
 &   0.27 &   0.14 &    0.10 &     0.09 \\
OREP.PA &   1458 &  14567 &   14689 &      146 \\
 &   0.27 &   0.19 &    0.18 &     0.09 \\
PERP.PA &    146 &   1458 &   14567 &      189 \\
 &   0.24 &   0.17 &    0.12 &     0.08 \\
PEUP.PA &    146 &   1458 &   14689 &    14567 \\
 &   0.27 &   0.17 &    0.12 &     0.11 \\
PRTP.PA &    146 &   1458 &   14689 &    14567 \\
 &   0.36 &   0.14 &    0.12 &     0.08 \\
PUBP.PA &    146 &   1458 &   14689 &     1246 \\
 &   0.19 &   0.16 &    0.15 &     0.09 \\
RENA.PA &    146 &  14689 &   14567 &  1456789 \\
 &   0.22 &   0.15 &    0.11 &     0.10 \\
SAF.PA  &    146 &  14689 &   14567 &      189 \\
  &   0.22 &   0.20 &    0.12 &     0.11 \\
SASY.PA &   1458 &  14567 &   14689 &    12458 \\
 &   0.15 &   0.14 &    0.12 &     0.10 \\
SCHN.PA &  14689 &   1458 &     146 &    14567 \\
 &   0.20 &   0.16 &    0.15 &     0.12 \\
SGEF.PA &  14689 &    146 &   14567 &     1458 \\
 &   0.21 &   0.18 &    0.15 &     0.12 \\
SGOB.PA &  14689 &    146 &     189 &    14567 \\
 &   0.19 &   0.17 &    0.14 &     0.11 \\
SOGN.PA &  14689 &    146 &   14567 &      189 \\
 &   0.21 &   0.18 &    0.10 &     0.09 \\
STM.PA  &    189 &    146 &    1458 &     1679 \\
  &   0.24 &   0.21 &    0.12 &     0.09 \\
TECF.PA &    146 &  14567 &   14689 &     1246 \\
 &   0.25 &   0.17 &    0.16 &     0.08 \\
TOTF.PA &  14689 &  14567 &     146 &   124689 \\
 &   0.24 &   0.16 &    0.15 &     0.09 \\
VIE.PA  &    146 &    189 &    1458 &    14689 \\
  &   0.32 &   0.12 &    0.12 &     0.12 \\
VIV.PA  &   1458 &  14689 &   14567 &      189 \\
  &   0.26 &   0.18 &    0.16 &     0.09 \\
VLLP.PA &    146 &    189 &   14567 &    14689 \\
 &   0.21 &   0.15 &    0.11 &     0.11 \\
VLOF.PA &    146 &  14567 &   14689 &      189 \\
 &   0.25 &   0.15 &    0.12 &     0.11 \\
\bottomrule
\end{tabular}

%% file: AIC-aggAModelRanking.tex
\begin{tabular}{rcccc}
\toprule
{} &      0 &      1 &      2 &        3 \\
\midrule
ACCP.PA &    146 &  14689 &    189 &  1456789 \\
 &   0.19 &   0.16 &   0.10 &     0.09 \\
AIRP.PA &   1679 &    146 &  14567 &    14689 \\
 &   0.31 &   0.14 &   0.14 &     0.12 \\
ALSO.PA &    146 &    189 &  14567 &     1246 \\
 &   0.21 &   0.17 &   0.12 &     0.08 \\
ALUA.PA &    146 &    189 &  14689 &    14567 \\
 &   0.25 &   0.14 &   0.12 &     0.10 \\
AXAF.PA &   1679 &  14689 &  14567 &      146 \\
 &   0.27 &   0.26 &   0.14 &     0.05 \\
BNPP.PA &  14689 &   1679 &    146 &    14567 \\
 &   0.31 &   0.15 &   0.08 &     0.08 \\
BOUY.PA &    146 &  14689 &  14567 &      189 \\
 &   0.23 &   0.16 &   0.12 &     0.09 \\
CAGR.PA &   1679 &    146 &  14567 &    14689 \\
 &   0.31 &   0.17 &   0.13 &     0.12 \\
CAPP.PA &    146 &  14689 &    189 &    14567 \\
 &   0.23 &   0.18 &   0.13 &     0.10 \\
CARR.PA &  14689 &    146 &   1679 &    14567 \\
 &   0.20 &   0.20 &   0.19 &     0.09 \\
DANO.PA &  14689 &   1679 &    146 &    14567 \\
 &   0.21 &   0.19 &   0.17 &     0.10 \\
EDF.PA  &    146 &  14689 &   1679 &     1246 \\
  &   0.21 &   0.15 &   0.14 &     0.10 \\
ESSI.PA &   1679 &  14567 &    146 &    14689 \\
 &   0.21 &   0.17 &   0.15 &     0.11 \\
LAGA.PA &    146 &    189 &   1289 &    14567 \\
 &   0.28 &   0.16 &   0.08 &     0.08 \\
LOIM.PA &    189 &    146 &  14689 &    14567 \\
 &   0.24 &   0.18 &   0.11 &     0.09 \\
LVMH.PA &   1679 &  14567 &  14689 &      146 \\
 &   0.29 &   0.22 &   0.16 &     0.08 \\
MICP.PA &    146 &  14689 &  14567 &     1246 \\
 &   0.23 &   0.16 &   0.11 &     0.08 \\
OREP.PA &   1679 &  14567 &  14689 &      189 \\
 &   0.21 &   0.19 &   0.17 &     0.12 \\
PERP.PA &   1679 &    146 &  14689 &    14567 \\
 &   0.19 &   0.17 &   0.14 &     0.13 \\
PEUP.PA &    146 &   1679 &  14689 &      189 \\
 &   0.26 &   0.19 &   0.13 &     0.07 \\
PRTP.PA &    146 &   1679 &  14689 &    14567 \\
 &   0.36 &   0.10 &   0.10 &     0.09 \\
PUBP.PA &    146 &   1679 &  14689 &      189 \\
 &   0.18 &   0.15 &   0.11 &     0.10 \\
RENA.PA &    146 &  14689 &    189 &  1456789 \\
 &   0.16 &   0.16 &   0.15 &     0.08 \\
SAF.PA  &    146 &  14689 &  14567 &     1246 \\
  &   0.23 &   0.19 &   0.12 &     0.09 \\
SASY.PA &   1679 &    146 &  14567 &    14689 \\
 &   0.16 &   0.14 &   0.13 &     0.13 \\
SCHN.PA &  14689 &    146 &   1679 &    14567 \\
 &   0.23 &   0.16 &   0.15 &     0.12 \\
SGEF.PA &    146 &   1679 &  14689 &    14567 \\
 &   0.16 &   0.15 &   0.15 &     0.14 \\
SGOB.PA &  14689 &    146 &    189 &     1246 \\
 &   0.18 &   0.18 &   0.10 &     0.08 \\
SOGN.PA &  14689 &    146 &    189 &     1246 \\
 &   0.24 &   0.19 &   0.14 &     0.09 \\
STM.PA  &    189 &    146 &   1679 &    14689 \\
  &   0.25 &   0.22 &   0.11 &     0.08 \\
TECF.PA &    146 &  14689 &  14567 &  1456789 \\
 &   0.21 &   0.20 &   0.15 &     0.09 \\
TOTF.PA &  14689 &  14567 &    146 &     1458 \\
 &   0.21 &   0.16 &   0.12 &     0.08 \\
VIE.PA  &    146 &   1679 &  14567 &    14689 \\
  &   0.21 &   0.15 &   0.14 &     0.13 \\
VIV.PA  &   1679 &  14689 &  14567 &   124689 \\
  &   0.27 &   0.25 &   0.14 &     0.07 \\
VLLP.PA &    146 &    189 &  14689 &     1458 \\
 &   0.24 &   0.21 &   0.12 &     0.10 \\
VLOF.PA &    146 &   1679 &    189 &    14567 \\
 &   0.19 &   0.14 &   0.12 &     0.12 \\
\bottomrule
\end{tabular}